\documentclass[11pt]{article}
\pdfoutput=1
\usepackage{amsmath}
\usepackage{amsthm}
\usepackage{amscd}
\usepackage{amsfonts}
\usepackage[psamsfonts]{amssymb}
\usepackage{graphicx}
\usepackage{hyperref}
\hypersetup{
    colorlinks=false,
    pdfborder={0 0 0},
}

\def\hri#1#2{\href{http://arxiv.org/abs/#1}{[#1 #2]}}
\def\hre#1{\href{http://arxiv.org/abs/#1}{[#1]}}

\numberwithin{equation}{section}

\makeatletter
\renewcommand\section{\@startsection {section}{1}{\z@}%
                                 {-3.5ex \@plus -1ex \@minus -.2ex}
                                   {2.3ex \@plus.2ex}%
                                   {\normalfont\large\bfseries}}
\renewcommand\subsection{\@startsection{subsection}{2}{\z@}%
                                   {-3.25ex\@plus -1ex \@minus -.2ex}%
                                     {1.5ex \@plus .2ex}%
                                     {\normalfont\bfseries}}
\renewcommand\subsubsection{\@startsection{subsubsection}{3}{\z@}%
                                   {-3.25ex\@plus -1ex \@minus -.2ex}%
                                     {1.5ex \@plus .2ex}%
                                     {\normalfont\itshape}}
\makeatother

\begin{document}

\begin{flushright}
\end{flushright}
\vspace{0.5in}
\begin{center}{\LARGE{\bf Holographic Renormalization  
} } \vspace{0.1in}
\\  {\LARGE{\bf of}} \\ \vspace{0.1in}
 {\LARGE{\bf Einstein-Maxwell-Dilaton Theories}}\\ 
\vspace{0.85in}

{\large \bf Bom Soo Kim}\\ ~\\ \vspace{-0.1in}
{\normalsize {Department of Physics and Astronomy,}} \\
{\normalsize {University of Kentucky, Lexington, KY 40506, USA}}\\ 
{\normalsize  \it  {bom.soo.kim@uky.edu}}\\ 
\end{center}
\bigskip 
\vspace{0.5in}

\abstract{
We generalize the boundary value problem with a mixed boundary condition that involves the gauge and scalar fields in the context of Einstein-Maxwell-Dilaton theories. In particular, the expectation value of the dual scalar operator can be a function of the expectation value of the current operator. The properties are prevalent in a fixed charge ensemble because the conserved charge is shared by both fields through the dilaton coupling, which is also responsible for non-Fermi liquid properties. We study the on-shell action and the stress energy tensor to note practical importances of the boundary value problem. In the presence of the scalar fields, physical quantities are not fully fixed due to the {\it finite boundary terms} that manifest in the massless scalar or the scalar with mass saturating the Breitenlohner-Freedman bound. 
}

\newpage
\tableofcontents
\newpage

\section{Introduction \& Salient Features} 

Einstein-Maxwell-Dilaton (EMD) theories are natural from the dimensional reductions of consistent string theory. They have provided various distinctive physical properties depending on the parameters present in the theory and have been studied extensively. See some earlier literature \cite{Gibbons:1987ps}-\cite{Holzhey:1991bx} and recent ones for condensed matter applications
\cite{Taylor:2008tg}-\cite{Gouteraux:2011ce} in the holographic context \cite{Maldacena:1997re}-\cite{Aharony:1999ti}. 

EMD theories have a $U(1)$ gauge field $A_\mu$ and a dilaton $\phi$ in addition to a metric $g_{\mu\nu}$, where $\mu, \nu$ run for all the coordinates.  Compared to the minimal coupling between the gauge field and a scalar, these theories have a scalar coupling with the form $W(\phi) F^2$ where $F=dA$ is the field strength. There is a conserved current $J^\mu$ that can be evaluated by the variation of the gauge field $A_\mu$.  
\begin{align} \label{Current1}
J^\mu \propto  \sqrt{-g} W(\phi) F^{r\mu}  \;, 
\end{align}
when the field $A_\mu (r)$ is only a function of a radial coordinate $r$. $g$ is the determinant of the metric. Upon a close examination, one finds the charge $J^0$ has contributions not only from the gauge field, but also from the scalar field through the function $W(\phi)$. Here we investigate the holographic renormalization of EMD theories with emphasis on the role of this coupling. We are going to focus on the theories with AdS asymptotics and analytic examples.  

The variation of the term $\sqrt{-g} W(\phi) F^2 $ in an action with respect to the gauge field $A_\nu$ provides a boundary contribution of the form 
\begin{align} \label{MaxwellTerm1}
\sqrt{-\gamma} W(\phi) ~n_r F^{r\nu} \delta A_\nu \;,
\end{align}   
where $\gamma$ is the determinant of a boundary metric, and $ n^r $ is a unit normal vector for a fixed radius. The notations are systematically explained below. It is consistent to impose the Dirichlet boundary condition $ \delta A_\nu =0 $ to have a well defined variational problem. This is the case of fixing the constant part of the gauge field, the chemical potential. Holographic renormalization of EMD theories with this Dirichlet boundary condition has been considered in \cite{Papadimitriou:2005ii}.  

Now let us consider an alternative quantization that sets $ \delta F^{r\nu} =0$. We are directed to add the boundary term 
\begin{align} \label{MaxwellTerm2}
\sqrt{-\gamma} W(\phi) ~n_r F^{r\nu} A_\nu \;.
\end{align}  
One can easily see a new feature for the variation of this term. We produce not only the variation $ \delta F^{r\nu} $, but also the scalar variation $\delta \phi$ because of $W(\phi)$. As we see below, it is natural to couple the variation of the gauge and scalar fields to have a consistent variational problem. Thus we seek a possibility to impose a generalized mixed boundary condition on both fields. The coupling term $\sqrt{-g} W(\phi) F^2$ brings forth this possibility naturally. 
The variation of \eqref{MaxwellTerm2} combined with \eqref{MaxwellTerm1} gives   
\begin{align} \label{VectorVariation1}
\sqrt{-\gamma} W(\phi) n_r A_\nu \left( \delta F^{r\nu} + 
 F^{r\nu} \frac{\partial \log W(\phi)}{\partial \phi}  \delta \phi \right) \;.
\end{align} 
The first term is expected, and one can impose the variational condition $ \delta F^{r\nu}=0$. As one already anticipated, there is an additional term. It can be simplified, using \eqref{Current1}, to 
\begin{align} 
(J^\nu A_\nu) \frac{\partial \log W(\phi)}{\partial \phi}  \delta \phi \;.
\end{align}
Note the combination $J^\nu A_\nu $ is usually finite at the boundary. 

This term prompts us to consider the variation of the gauge field together with that of the scalar. The canonically normalized scalar kinetic term, upon variation, gives the boundary contribution $-\sqrt{-\gamma} n^r \partial_r \phi (\delta \phi)$. We consider two possible boundary terms for the scalar field with appropriate coefficients following the previously developed variational approaches \cite{Marolf:2006nd} 
\begin{align} \label{TwoScalarTerms2}
\sqrt{-\gamma} \left[ \frac{\Lambda_\phi}{2L} \phi^2 + c_\phi \phi n^r \partial_r \phi \right] \;. 
\end{align}
Here we leave the coefficients $\Lambda_\phi, c_\phi$ unfixed. Then putting the variations together, we get  
\begin{align} 
&\sqrt{-\gamma} \left\{ \big[ (c_\phi -1) n^r \partial_r \phi + \frac{\Lambda_\phi}{L} \phi + 4 \frac{\partial W}{\partial \phi} ~n_r F^{r\nu} A_\nu \big] (\delta \phi)  \right. \\
&\qquad\qquad \bigg. + c_\phi \phi n^r \delta(\partial_r \phi)  + 4 W n_r  A_\nu (\delta F^{r\nu}) \bigg\}  \;. \nonumber
\end{align}
Note the term $\frac{\partial W}{\partial \phi} ~n_r F^{r\nu} A_\nu (\delta \phi)$ mixes the scalar variation with the gauge field dependent term and plays a crucial role.  \\

Investigating this variational problem in more general context is the main task of the paper. The basics of the variational problem are described in two sections, \S \ref{sec:SeparateVariation} and \S \ref{sec:CoupledVariation}, with some review. The program is carried out systematically for two different forms of the coupling $W(\phi)$, the exponential coupling $W(\phi) \sim e^\phi $ in \S \ref{sec:BoundaryProblemExponentialW} (with two examples in \S \ref{sec:EMDAdS4} and \S \ref{sec:EMDAdS5}) and the polynomial coupling $W(\phi) \sim \phi^k $ in \S \ref{sec:BoundaryProblemPolynomialW} (with an example in \S \ref{sec:Interpolating}). 
We contrast the boundary value problem of the EMD theories with that of the theories with a  minimal coupling in \S \ref{sec:MinimalCoupling}. 

In parallel, we also carefully examine the on-shell action and the stress energy tensor of the EMD theories. The basics along with some review are presented in the beginning of \S \ref{sec:General} and in \S \ref{sec:OnShellAction}. They are applied to the three examples in \S \ref{sec:EMDSolutions}. It is pleasant to see that the results of the general variational problem actually fit together nicely with the analysis of the on-shell action and the stress energy tensor.

Along the way, our investigations direct us to appreciate two physical implications. One is the finite boundary or counter terms considered in \S \ref{sec:FiniteCounterTerms}. There we attempt to compare the holographic finite boundary terms to the similar notion of finite radiative corrections in Quantum field theory {\it \`a la} Jackiw. We also try to survey earlier literature with finite counter terms in holography. Another implication is non-Fermi liquid properties and splitting of the conserved charge due to the dilaton coupling. It is presented in \S \ref{sec:HiddenCharge}. 

Before moving on, let us list some lessens we have learned with this investigation.

\begin{itemize}
\item In the context of EMD theories, the boundary variational problem can be generalized to include the mixed boundary condition between the gauge and scalar fields.

\item For the fixed charge ensemble, the expectation value of the dual scalar operator $\langle \mathcal O_{\alpha} \rangle $ can be a function of the expectation value of the dual current operator $\langle \mathcal O_{Q_F} \rangle$  
\begin{align}
\langle \mathcal O_{\alpha} \rangle =  c~   Q_F  \langle \mathcal O_{Q_F} \rangle  + \cdots \;,
\end{align}
with some additional contributions in general. There are some conditions on $\alpha$ and $Q_F$ at the boundary. $c$ is a constant. 
 
\item The general boundary value problem can impose conditions on the parameters, $\Lambda_\phi$ and $c_\phi$, of the boundary terms. This can be different from the condition that renders the on-shell action and the stress energy tensor finite. This happens when the boundary terms provide finite contributions. Then the on-shell action and the stress energy tensor depend on the parameters. The condition obtained from the variation problem can be used to render the mass evaluated by the stress energy tensor to ADM mass. Furthermore, the differential form of the first law of thermodynamics is satisfied. 

\item Our variational problem reveals that the finite boundary terms, whose coefficients are not fixed by the requirements of the theory, are general features of the theories with the scalar fields. This is especially clear for the massless scalar and for the scalar with mass saturating the BF bound when their solutions are realized with the faster falloff at the boundary. 

\item  The dilaton coupling provides a way to share the conserved charge between the gauge and scalar fields through the coupling. The physical properties due to this coupling can be clarified with examples. The first two examples in \S \ref{sec:EMDSolutions} have $W \sim 1/4$ at the boundary, and thus effectively the entire charge comes from the gauge field. The other example gives a non-trivial boundary profile for $W$. The conserved charge is shared by the scalar, and the system exhibits a non-trivial and interesting physics such as a non-Fermi liquid state. 

\end{itemize} 
These lessens are summarized in conclusion \S \ref{sec:Conclusion} with two illustrations, one for the variational problem and another for the on-shell action.

\section{Generalities} \label{sec:General}

We consider the Einstein-Maxwell-Dilaton theory in $d+1$ dimensional asymptotically AdS space  
\begin{align}  \label{EMDAction}
\begin{split}
S &=\frac{1}{2\kappa^2} \int d^{d+1} x \sqrt{-g} \left[R  -W(\phi) F^2 -\frac{1}{2} (\partial \phi)^2 - V(\phi) \right] \;, \\
ds^2 &= \left[ \frac{r^2}{L^2} + \cdots \right] \left(-h_1(r) dt^2 + d\vec x^2  \right) + \left[ \frac{r^2}{L^2} + \cdots \right]^{-1} \frac{ dr^2}{h_2(r)}  \;,
\end{split}
\end{align}  
where $\kappa^2 =8\pi G$, and $ h_1(r)= 1 + \cdots, h_2(r)= 1 + \cdots$, where $\cdots$ are possible sub-leading terms in large $r$ expansion.   

The Einstein, Maxwell and scalar equations are 
\begin{align}\label{EMDEOM}
\begin{split}
&R_{\mu\nu} -\frac{1}{2} R g_{\mu\nu} = \frac{1}{2} \partial_\mu \phi \partial_\nu\phi +2 W F_{\mu\rho} F^{~\rho}_{\nu}
- g_{\mu\nu}\left( \frac{1}{4} (\partial \phi)^2 + \frac{W}{2} F^2+  \frac{1}{2}  V(\phi) \right) \;, \\
&\partial_\mu \left( \sqrt{-g} W F^{\mu\nu} \right) = 0 \;,  \\
&\frac{1}{\sqrt{-g}} \partial_\mu \left( \sqrt{-g} g^{\mu\nu} \partial_\nu \phi \right) -\frac{\partial W}{\partial \phi} F^2 -\frac{\partial V}{\partial \phi}=0\;. 
\end{split}
\end{align} 
A particular solution depends on the form of the gauge coupling $W(\phi)$. We consider the gauge and scalar fields depending only on the radial coordinate, $F=dA, A=A_t(r) dt$ and $\phi(r)$. Then the second equation gives a finite and constant value  that can be identified as a charge
\begin{align}\label{ConservedMaxwellCharge}
q = -\frac{4}{2\kappa^2} \sqrt{-g} W F^{rt} \;. 
\end{align}
 
In two sections \S \ref{sec:SeparateVariation} and \S \ref{sec:CoupledVariation}, we investigate the general boundary variational problem of the EMD theories. With the results of the two sections, we carefully revisit the on-shell action and the stress energy tensor in \S \ref{sec:OnShellAction}. They are contrasted to the boundary value problem of the theories with a minimal coupling in \S \ref{sec:MinimalCoupling}.

\subsection{Scalar or vector variation} \label{sec:SeparateVariation}

Holographic renormalization is a central part of extracting physical quantities of the dual field theory from the gravity side \cite{Henningson:1998gx}-\cite{Ishibashi:2004wx}. See also a previous work in the context of EMD theories \cite{Papadimitriou:2005ii} that considered the scalar and vector variations separately. We examine the possible boundary terms, the on-shell action, and stress energy tensor. 

The variational principle for scalar and Maxwell fields have been studied in \cite{Witten:2001ua}\cite{Berkooz:2002ug}\cite{Sever:2002fk}\cite{Marolf:2006nd}. Here we review them following \cite{Marolf:2006nd} to have a fresh look on the problems. The variation of the canonically conjugate scalar kinetic term contains the boundary contribution
\begin{align} \label{ScalarBContribution}
\delta S_\phi = -\frac{1}{2\kappa^2} \int d^d x \sqrt{-\gamma} ~n^r \partial_r \phi (\delta \phi) \;. 
\end{align}
To satisfy the equation of motion, it is required to impose a boundary condition for the scalar. For this purpose, we introduce a general boundary term 
\begin{align} \label{ScalarBTerm2}
S_{b}(\phi) =  \frac{1}{2\kappa^2} \int d^d x \sqrt{-\gamma} \left(c_\phi \phi n^r \partial_r \phi + \frac{\Lambda_\phi}{2L} \phi^2 \right) \;.
\end{align} 
Note that we introduce the coefficients $\Lambda_\phi$ and $c_\phi$ unfixed. Both terms were considered in \cite{Marolf:2006nd} mostly with some particular values of $\Lambda_\phi$ and $c_\phi$, while the general mixed boundary conditions were also advertised.  

The boundary behavior of the scalar 
\begin{align} \label{GeneralScalarAsymptotics}
\phi \to \frac{\alpha(x)}{r^{\lambda_-}} + \frac{\beta(x)}{r^{\lambda_+}} \;, \qquad 
\lambda_\pm = \frac{d}{2} \pm \frac{1}{2} \sqrt{d^2 + 4 m^2_\phi L^2} \;,
\end{align} 
depends on its mass
\begin{align}
m^2_\phi L^2 = L^2 \frac{\partial^2}{\partial \phi^2} \left(  V(\phi) +  W(\phi)  F^2 \right)\bigg|_{\phi=0}  \;.
\end{align}  
To have a well defined variational problem, one needs to impose a boundary condition. There are three different cases with three different mass ranges, $m_\phi^2 L^2 \geq 1-d^2/4 $, $ -d^2/4 < m_\phi^2 L^2 < 1-d^2/4 $, and $ m_\phi^2 L^2 = -d^2/4  $. The last one saturates the Breitenlohner-Freedman (BF) bound \cite{BF1}\cite{BF2}. For the mass range slightly above the BF bound, $ -d^2/4 < m_\phi^2 L^2 < 1-d^2/4 $, the two falloffs are both normalizable \cite{Klebanov:1999tb}, and it is possible to impose boundary condition on $\alpha, \beta$ or on a linear combination of them \cite{Witten:2001ua}. See also previous discussions on the generalized boundary conditions on the scalar variations \cite{Henneaux:2002wm}-\cite{Anabalon:2015xvl}. 

Here we illustrate the variational problem with the scalar mass saturating the BF bound, $m_\phi^2 L^2 =-d^2/4 $. The scalar behaves as 
\begin{align} \label{ScalarFalloffBFBound}
\phi \to \frac{\alpha(x) \log r}{r^{d/2}} + \frac{\beta(x)}{r^{d/2}}  \;.
\end{align} 
Consider the boundary contribution \eqref{ScalarBContribution} and the variation of boundary term \eqref{ScalarBTerm2} for special case of $c_\phi=0$ 
\begin{align}
\delta S_\phi + \delta S_{b}(\phi) = \int d^d x \frac{(\delta \alpha \log r + \delta \beta)}{2\kappa^2 L^{d+1}}  
\left[ \left( \Lambda_\phi + \frac{d}{2} \right) (\alpha \log r + \beta) -\alpha  \right] \;.
\end{align}
Here setting $\Lambda_\phi = -d/2$ and $\alpha=0$ provides a well defined variational problem.  

On the other hand, for $c_\phi=1$, the variation yields 
\begin{align} 
\delta S_\phi + \delta S_{b}(\phi) = \int d^d x  \frac{(\alpha \log r + \beta) }{\tilde \kappa^2 }  \left[\delta \alpha \left(1 + \big(\Lambda_\phi - \frac{d}{2} \big) \log r \right) + \delta \beta \big(\Lambda_\phi - \frac{d}{2} \big)  \right] \;.
\end{align}
Here we define $\tilde \kappa^2 =2\kappa^2 L^{d+1} $.
There are two different ways to have a well defined variational problem. First it works for $\Lambda_\phi = d/2$ followed by setting $\alpha=0$. Then the expectation value (density for both space and time) of the dual scalar operator is given by 
\begin{align}
\langle \mathcal O _{\alpha=0} \rangle = \frac{\beta}{\tilde \kappa^2} \;.
\end{align}
Second, we consider to impose the condition with a standard $\log r$ dependence
\begin{align}
\Lambda_\phi =  \frac{d}{2} - \frac{1}{\log r} \;. 
\end{align} 
The resulting variation is  
\begin{align}
\delta S_\phi + \delta S_{b}(\phi) = -\int d^d x  \frac{\alpha  \delta  \beta}{\tilde \kappa^2 } \;.
\end{align}  
Then the variational principle is well defined for fixed $\beta$. And the corresponding scalar expectation value is 
\begin{align}
\langle \mathcal O _{\beta} \rangle = - \frac{\alpha}{\tilde \kappa^2} \;.
\end{align}
There are more than one possible quantization for different values of the parameter $\Lambda_\phi$. Below we seek more general possibilities by utilizing the parameters $\Lambda_\phi$ and $c_\phi$. \\

The variational problem for the gauge field is also closely examined in \cite{Marolf:2006nd}. For the Maxwell action with $W=1/4$, the condition \eqref{ConservedMaxwellCharge} enables us to read off the sub-leading contribution for $A_t$ at the boundary 
\begin{align}
A_t  = \mu + \frac{2 q \kappa^2}{2-d} \frac{L^{d-1}}{r^{d-2}} + \cdots \;, \qquad \text{for} \quad W(\phi)=\frac{1}{4}\;.
\end{align} 
Where $\cdots$ are sub-leading contributions in the large $r$ expansion. The variation of the action with respect to the gauge field $A$ produces the boundary contribution 
\begin{align} \label{MaxwellVariation}
 -\frac{1}{2\kappa^2} \int d^d x \sqrt{-\gamma}  ~n_r F^{r\nu} (\delta A_\nu) \;. 
\end{align}
Once $\delta A_\nu =0$ is imposed at the boundary, the variational problem is well defined. This fixes the chemical potential $\mu$ of the time component $A_t$. There is an alternative quantization, fixing the charge, that can be done by adding the boundary term 
\begin{align} \label{MaxwellBoundaryTerm}
\frac{1}{2\kappa^2} \int d^d x \sqrt{-\gamma} ~n_r F^{r\nu} A_\nu \;. 
\end{align} 
Then the variation \eqref{MaxwellBoundaryTerm} combined with \eqref{MaxwellVariation} shows that the variational problem is well defined with the condition $\delta F^{r\nu} = 0 $. This fixes the coefficient of the sub-leading contribution of $A_t$, which is the electric charge.

\subsection{Scalar and Vector variations : coupled} \label{sec:CoupledVariation}

Let us examine the Maxwell term of the EMD action. The variation of the action with respect to the gauge field $A$ produces the boundary contribution
\begin{align} \label{EMDMaxwellBoundaryContribution}
\delta S_A = -\frac{2}{\kappa^2} \int d^d x \sqrt{-\gamma} W(\phi) ~n_r F^{r\nu} (\delta A_\nu) \;.
\end{align}
Again imposing $\delta A_\nu =0$ at the boundary renders the variational problem well defined, and it amounts to fix the chemical potential $\mu$ of $A_t$. 

Do the EMD theories have an analogous quantization of fixing charge? To answer the question, let us add the following boundary term 
\begin{align} \label{EMDMaxwellBoundaryTerm}
S_{b}(A) = \frac{2}{\kappa^2} \int d^d x \sqrt{-\gamma} W(\phi) ~n_r F^{r\nu} A_\nu \;,
\end{align}
which is a direct generalization of the \eqref{MaxwellBoundaryTerm}. The boundary term $-\frac{4}{2\kappa^2} \sqrt{-\gamma} ~W n_r F^{rt} A_t = \mu q$ is actually finite. By combining the \eqref{EMDMaxwellBoundaryContribution} and the variation of \eqref{EMDMaxwellBoundaryTerm}, we get 
\begin{align}
\frac{2}{\kappa^2} \int d^d x \sqrt{-\gamma} W(\phi) F^{r\nu} n_r A_\nu  \left( \frac{\delta F^{rt}}{F^{rt}} + \frac{\partial \log W(\phi)}{\partial \phi}  \delta \phi \right)  \;.
\end{align} 
In general, the alternative quantization for $\delta A_\nu$ in EMD theories is not just fixing the sub-leading boundary contribution of the gauge field $F_{rt} = \partial_r A_t $. This allows more freedom for the boundary condition for the gauge field. The combination 
\begin{align}
\frac{\delta F^{rt}}{F^{rt}} + \frac{\partial \log W(\phi)}{\partial \phi}  \delta \phi 
\end{align}
needs to be fixed. The boundary condition depends on the details of the solution $W(\phi)$, $\phi$ and $F^{rt}$ at the boundary. Note that the variation of gauge field is tied to the variation of the scalar. As we see below, it is natural to fix a mixed boundary condition between the gauge and scalar fields. 

Thus we consider general boundary contributions and boundary terms.  
\begin{align} \label{GeneralBoundaryTerms}
&\delta S_M =\delta S_\phi + \delta S_A + \delta S_b(\phi) + \delta S_b (A) \nonumber \\
&=\frac{1}{2\kappa^2} \int d^d x \sqrt{-\gamma} \left[ -n^r \partial_r \phi (\delta \phi)- 4 W n_r F^{r\nu} (\delta A_\nu)+ \frac{\Lambda_\phi}{2L} \delta (\phi^2) + c_\phi \delta(\phi n^r \partial_r \phi )  + 4 c_F \delta( W n_r  F^{r\nu} A_\nu ) \right] \nonumber  \\
&=\frac{1}{2\kappa^2} \int d^d x \sqrt{-\gamma} \left\{ \left[ (c_\phi -1) n^r \partial_r \phi + \frac{\Lambda_\phi}{L} \phi + 4 c_F \frac{\partial W}{\partial \phi} ~n_r F^{r\nu} A_\nu \right] (\delta \phi)+ c_\phi \phi n^r \delta(\partial_r \phi)   \right. \\
&\qquad\qquad\qquad\qquad\qquad\qquad  \bigg. + 4 (c_F-1) W n_r  F^{r\nu} (\delta A_\nu) + 4 c_F W n_r  A_\nu (\delta F^{r\nu}) \bigg\}  \;. \nonumber
\end{align}
If one considers $c_F=0$, the gauge and scalar variations are separated, and one already considered the case above. This case is usually corresponding to fix the chemical potential, and is referred as grand canonical ensemble. 

For EMD theories with $c_F=1$, we get 
\begin{align} \label{EMDBTDFOne}
\begin{split}
\delta S_M^{s-can} = \int d^d x \frac{\sqrt{-\gamma}}{2\kappa^2} &\left\{ \left[ (c_\phi -1) n^r \partial_r \phi + \frac{\Lambda_\phi}{L} \phi + 4 \frac{\partial W}{\partial \phi} ~n_r F^{r\nu} A_\nu \right] (\delta \phi) \right. \\
&\qquad\qquad  \bigg. + c_\phi \phi n^r \delta(\partial_r \phi) + 4 W n_r  A_\nu (\delta F^{r\nu}) \bigg\}  \;. 
\end{split}
\end{align}
Here we note the mixing term $4 \frac{\partial W}{\partial \phi} ~n_r F^{r\nu} A_\nu  (\delta \phi) $. If $W$ is independent of the scalar field $\phi$, it correspond to fix the charge of the gauge field and to a canonical ensemble. EMD theories are different. We call it {\it semi canonical} ensemble. 
 
We consider both the grand and semi canonical cases by using \eqref{GeneralBoundaryTerms}. The boundary condition depends on the form of the coupling $W(\phi)$, and thus we examine two different classes separately, $W(\phi) \sim e^{\phi }$ and $W(\phi) \sim \phi^k$, in \S \ref{sec:CouplingW}. For the rest of the section, we lay out the general formulas for the on-shell action and stress energy tensors.

\subsection{On-shell action \& Stress energy tensor} \label{sec:OnShellAction}

Let us introduce some notations following Brown and York \cite{Brown:1992br}. Our formalism utilize two kinds of hyper-surfaces, the time-like boundary surface at a large fixed $r$ and the space-like surface at a fixed time $x^0$. The projections onto these hyper-surfaces require two normal vectors to the surfaces. The ADM form of the metric for the $d$ dimensional time-like hyper-surface homomorphic to boundary $\partial M$ has the form 
\begin{align} \label{ADMForm}
ds^2 = N_r^2 dr^2 + \gamma_{ij} (dx^i + N^i dr)(dx^j + N^j dr) \;,
\end{align}
where $x^i$ ($i=0,1, \cdots, d-1$) are the coordinates spanning a given time-like surface, while $r$ is the holographic coordinate. The corresponding unit normal vector is 
\begin{align}
n_\mu = N_r (0,0, \cdots ,0,1) \;,
\end{align}
where the components are ordered as $(x^i, r)$. We also define the time-like unit normal of a space-like surface by
\begin{align} 
u_\mu = -N_\Sigma (1,0,\cdots ,0,0) \;,
\end{align}
where $ u_\mu$ defines the local flow of time in $\partial M$, and $N_\Sigma$ comes from another ADM decomposition
\begin{align} \label{ADMForm2}
ds_\gamma^2 =& \gamma_{ij} dx^i dx^j 
=  -N_\Sigma^2 dx^0 dx^0 + \sigma_{ab} (dx^a + N_\Sigma^a dx^0)( dx^b + N_\Sigma^b dx^0) \;.
\end{align}
Where $a=1, \cdots, d-1$, spanning the spatial coordinates at the boundary $\partial M$.

The projections onto the $d$-dimensional time-like boundary hyper-surface and the $(d-1)$-dimensional space-like intersection surface are given by
\begin{align}
\gamma_{\mu\nu} = g_{\mu\nu} - n_\mu n_\nu \;, \qquad  \qquad 
\sigma_{\mu\nu} = g_{\mu\nu} - n_\mu n_\nu + u_\mu u_\nu \;.
\end{align}
Since they are projection operators, they do not have inverses and the $(d+1)$-dimensional indices are raised and lowered by the metric $g_{\mu\nu}$. However, if we restrict them to the appropriate components, for example, $\gamma_{ij}$ with $i,j $ on the time-like hyper-surface, they have well-defined inverses and can be defined as the metric on the surface.

We evaluate the corresponding on-shell action. The first equation of \eqref{EMDEOM} gives 
\begin{align}
R= \frac{d+1}{d-1} V(\phi) + \frac{1}{2} (\partial \phi )^2 + \frac{d-3}{d-1} W F^2\;.
\end{align}
Then we have the on-shell action 
\begin{align} \label{GeneralOnShellAction}
S_{on-shell} &=\frac{1}{2\kappa^2} \int d^d x \int_{r_0}^{r_\epsilon} dr \sqrt{-g} ~\frac{2}{d-1} \left[V -WF^2  \right] + S_{b} \;, 
\end{align}
where $r_\epsilon$ is the UV boundary cut-off, and $r_0$ can be a horizon in the presence of black hole or $0$ for a zero temperature background. $S_b$ has possible boundary terms that make the variational principle work and counter terms that yield the action finite. It contains the well known Gibbons-Hawking \cite{Gibbons:1976ue} and Balasubramanian-Kraus counter terms \cite{Balasubramanian:1999re}
\begin{align} \label{GeneralboundaryTerms11}
S_{b} &= - \frac{1}{\kappa^2} \int_{r=r_\epsilon } d^d x \sqrt{-\gamma} \left[ \Theta + \frac{d-1}{L} + \frac{L}{2(d-2)} R^{d}\right] + S_{b}(A) + S_{b}(\phi) \;. 
\end{align}
Here $\gamma$ and $R^{d}$ are the metric at fixed $r$, and the scalar curvature of the metric $\gamma$. The Gibbon-Hawking term is the trace of $\Theta_{ij}$, the extrinsic curvature, defined as
\begin{align}
\Theta_{ij} = {\gamma_i}^\mu{\gamma_j}^\nu \Theta_{\mu\nu} \;, \qquad 
\Theta_{\mu\nu} = -\frac{1}{2} ( \nabla_\mu n_\nu + \nabla_\nu n_\mu )  \;.
\end{align}

The last two terms in \eqref{GeneralboundaryTerms11}, $S_{b}(A)$ and $ S_{b}(\phi)$, are possible boundary terms for the gauge and scalar fields introduced in the previous section.  
Collecting all the boundary terms that are used for the variation \eqref{GeneralBoundaryTerms}, we get  
\begin{align} \label{FullOnShellAction}
S_{on-shell} &=\frac{1}{2\kappa^2} \int d^d x \left\{ \int_{r_0}^{r_\epsilon} dr \sqrt{-g} ~\frac{2}{d-1} \left[V -WF^2  \right] \right. \\ 
&\qquad  \left. -2 \sqrt{-\gamma} 
\left[ \Theta + \frac{d-1}{L} - \frac{L\mathcal R^{(d)}}{2(d-2)}  -\frac{c_\phi}{2} \phi n^r \partial_r \phi -\frac{\Lambda_\phi}{4 L} \phi^2 
- 2 c_F W n_r F^{rt} A_t \right] \right\} \;. \nonumber 
\end{align}
This on-shell action is one central object we consider.

Another key objects are the Brown-York conserved quantities such as mass (energy), pressure, and angular momentum, which can be worked out from the stress energy tensor. Once one has the full on-shell action \eqref{FullOnShellAction}, one can compute the corresponding stress energy tensor \cite{Brown:1992br}\cite{Balasubramanian:1999re}. It is given by 
\begin{align}
\kappa^2 ~T_{ij} =  \Theta_{ij} - \Theta \gamma_{ij} - \frac{2}{\sqrt{-\gamma}} \frac{\delta \mathcal L_b}{\delta \gamma^{ij}}  \;,
\end{align}
where $\mathcal L_b $ is the Lagrangian for the boundary action $S_b$ without the Gibbons-Hawking term. 
Explicitly for \eqref{FullOnShellAction}
\begin{align} \label{FullStressEnergyTensor}
\kappa^2 T_{ij} &= \Theta_{ij} - \Theta \gamma_{ij} - \frac{d-1}{L} \gamma_{ij}
- \frac{L}{d-2} G^{(3)}_{ij} +\gamma_{ij} \left[\frac{c_\phi}{2} \phi n^r \partial_r \phi+ \frac{\Lambda_\phi}{4 L} \phi^2 \right] \\
&\qquad\qquad+ c_F \Big[ 2 \gamma_{ij} W n_r F^{rt} A_t -4 W n^r F_{r i} A_{j}  \Big] \;. \nonumber 
\end{align}
Note that the first four terms on the right hand side are from the Gibbons-Hawking term and the counter terms for AdS \cite{Brown:1992br}\cite{Balasubramanian:1999re}. The terms with $c_\phi$ and $\Lambda_\phi$ has been also considered in the context of the variational problem in \cite{Marolf:2006nd} for particular fixed coefficients.   

Given the stress energy tensor, we can proceed to compute a conserved charge associated with a killing vector $\xi^i$ that generates an isometry of the boundary geometry as in \cite{Brown:1992br}
\begin{align}
Q_\xi = \int_\Sigma d^{d-1} x \sqrt{\sigma} (u^i T_{ij} \xi^j) \;.
\end{align}
In particular, the mass density $M$ and trace of pressure $P$ can be computed solely from the stress energy tensor $T_{ij}$ and metric as
\begin{align} \label{MassPressure} 
\begin{split}
&\mathcal M = \int d^{d-1} x  ~M = \int d^{d-1} x \sqrt{\sigma} N_\Sigma u_i u_j T^{ij} \;, \\
&P = \frac{1}{d-1} \sigma^{ab} s_{ab} = \frac{1}{d-1} \sigma^{ab} \sigma_{ai} \sigma_{bj} T^{ij} \;. 
\end{split}
\end{align}
The mass density $M$ is identified as the time component of the stress energy tensor  $\langle T_{00} \rangle$ of the field theory.

There is an equivalent way to extract the energy and pressure. To do so, we convert the stress energy tensor $T_{ij}$ to the field theory stress energy tensor $ \langle T_{ij} \rangle  $.     
Let us identify the metric $\tilde g$ of the field theory, which can be read off at $r \to \infty$ from the metric in \eqref{EMDAction}    
\begin{align} 
ds^2 &\quad \overrightarrow{\text{\tiny $r \to \infty$}} \quad  \frac{r^2}{L^2}  \left(- dt^2 + d\vec x^2  \right) + \frac{L^2}{r^2} dr^2  
\equiv  \frac{r^2}{L^2} \tilde g_{ij} dx^i dx^j + \frac{L^2}{r^2} dr^2 \;,
\end{align} 
where $i, j = 0, 1, \cdots , d-1$ are the coordinates of the field theory and $a,b= 1, 2, \cdots, d-1$ are the spatial ones, while $\mu, \nu$ includes the radial coordinate as well.  
The field theory stress energy tensor $ \langle T^{ij} \rangle  $ can be computed using the relation \cite{Myers:1999psa} (also used recently in \cite{Dong:2012se}\cite{Kim:2012nb})
\begin{align}
\sqrt{-\tilde g} \tilde g_{ik} \langle T^{kj} \rangle 
= \lim_{r \to \infty} \sqrt{-\gamma} \gamma_{ik} T^{kj} \;.
\end{align}
We note that the field theory metric is flat. With this we get 
\begin{align}
\langle T_{ij} \rangle = \frac{r^{d-2}}{L^{d-2}} T_{ij} \;.
\end{align}
The indices are raised or lowered by the metric $\tilde g$. $\langle T_{00} \rangle$ and $\langle T_{aa} \rangle$ are the energy density and pressure of the field theory side. They coincide with the expressions \eqref{MassPressure}, which were evaluated using the stress energy tensor $T_{ij}$. 
\begin{align} \label{MassPressureEquivalence} 
&\langle T_{00} \rangle = E = \sqrt{\sigma} N_\Sigma u_i u_j T^{ij} \;, \qquad
\langle T_{aa} \rangle = P =  \sigma_{ai} \sigma_{bj} T^{ij} \;.
\end{align} 
We have checked this equivalence for the examples we considered below and also other examples in \cite{Brown:1994gs}, where the equivalence is established as a function of a finite radius $r=R$.

\subsection{Comparison to minimal coupling}  \label{sec:MinimalCoupling} 

From above, we have learned the boundary value problem is modified due to the dilaton coupling. Especially the variation of the scalar field has an additional term proportional to the charge. 

Will there be a qualitative difference for the minimal coupling in $d+1$ dimensions? Let us consider the action 
\begin{align}  \label{MaxwellActionMinimalCoupling}
S &=\frac{1}{2\kappa^2} \int d^{d+1} x \sqrt{-g} \left[R  -\frac{1}{4} F^2 -\frac{1}{2} (D \phi)^2 - V(\phi) \right] \;, 
\end{align}    
where $D_\mu = \partial_\mu - i A_\mu $. 
We focus on the solution with $\phi(r)$ and $F=dA, A=A_t(r) dt$. 
The boundary terms for the matter are 
\begin{align} \label{MinimalBoundaryContributions}
\delta S_{minimal} = -\frac{1}{2\kappa^2} \int d^d x \sqrt{-\gamma} ~\Big[ n_r F^{r \nu} (\delta A_\nu) + n^r (\partial_r \phi - iA_r \phi)(\delta \phi) \Big] \;. 
\end{align} 
Is there a coupling between the gauge field and scalar at the boundary? The term 
\begin{align}
- i n^r A_r ~\phi (\delta \phi)  \qquad\qquad : \quad \text{pure gauge} 
\end{align}
does not have physical effects because $A_r$ can be gauged away. Thus we confirm that there is no direct coupling between the gauge and scalar fields at the boundary.

\section{Dilaton Coupling $W(\phi)$} \label{sec:CouplingW}

In this section we perform a detailed analysis on the variational problem with some general forms of the Dilaton coupling $W(\phi)$. We consider in detail the exponential coupling $W(\phi) \sim e^{\phi} $ and the polynomial coupling $W(\phi) \sim \phi^k $. 

\subsection{Exponential coupling: $W(\phi) \sim e^{\phi}$} \label{sec:BoundaryProblemExponentialW}

We consider the exponential coupling $W(\phi) = \frac{1}{4} e^{c_W \phi}$. Then, $\frac{\partial W(\phi)}{\partial \phi} = c_W W (\phi)$. The new term in \eqref{GeneralBoundaryTerms}, after using \eqref{ConservedMaxwellCharge}, becomes 
\begin{align} \label{NewTermExponentialCoupling}
c_F \frac{4}{2\kappa^2} \sqrt{-\gamma}  \frac{\partial W}{\partial \phi} ~n_r F^{r\nu} A_\nu (\delta \phi) = c_F c_W \mu q (\delta \phi) \;.
\end{align}
It is independent of the details of the gauge field except the dependence of $c_W$ and $\mu q$. 
Here we consider several different cases for the scalar mass, including the mass slightly above the BF bound with two different normalizable modes, the mass saturating the BF bound, and the massless scalar in turn.   

\subsubsection{Scalar mass above the BF bound} \label{sec:ExponentialWAboveBFBound}

For the scalar mass slightly above the BF bound, we use \eqref{GeneralScalarAsymptotics}. For general $\Lambda_\phi$ and $c_\phi$, the variation \eqref{GeneralBoundaryTerms} has the form 
\begin{align}  \label{VarExpGen}
\begin{split}
\delta S_M 
&= \int d^d x \left\{ 
\frac{(\Lambda_\phi + [1-2c_\phi] \lambda_-) \alpha \delta \alpha(x)}{\tilde \kappa^2 r^{-d+2\lambda_-}} \right. \\
&\qquad\qquad+\frac{(\Lambda_\phi - c_\phi \lambda_- - [c_\phi-1] \lambda_+ ) \beta \delta\alpha + (\Lambda_\phi - c_\phi \lambda_+ - [c_\phi-1] \lambda_- ) \alpha \delta\beta(x)}{\tilde \kappa^2 } \\
& \left. \qquad\qquad+ (c_F -1) \mu q \frac{\delta \mu(x)}{\mu} + c_F \mu q \frac{\delta Q_F(x)}{Q_F}\right\} \;,
\end{split}
\end{align} 
where we abbreviate $\tilde \kappa^2 = 2\kappa^2 L^{d+1} $, and use the form for the field strength 
\begin{align} \label{FieldStrengthBoundaryQ}
F_{rt} = \frac{Q_F(x)}{r^{\lambda_Q}} \;. 
\end{align}
Note that $q$ and $Q_F$ can be, in general, different due to the non-trivial coupling $W$. This is more clear in the following section \S \ref{sec:BoundaryProblemPolynomialW}. The term \eqref{NewTermExponentialCoupling} that couples the charge and the scalar variation decays at least $ r^{-\lambda_-}$ and does not contribute. Thus the case at hand is effectively the same as the variational problem for the constant coupling $W$. Hereafter we suppress the coordinate dependence of the variations for simplicity. 

Here we impose 
\begin{align} \label{ConditionForAboveBFBoundExponentialW} 
\Lambda_\phi + [1-2c_\phi] \lambda_- = 0 
\end{align}
to have a well defined variational problem. This also renders the on-shell action finite. Then 
\begin{align} 
\delta S_M 
&\propto~ (\lambda_- -\lambda_+ ) \frac{[c_\phi-1]  \beta \delta\alpha +  c_\phi  \alpha \delta\beta}{\tilde \kappa^2 } + (c_F -1) \mu q \frac{\delta \mu}{\mu} + c_F \mu q \frac{\delta Q_F}{Q_F} =0 \;.
\end{align} 
Now it is clear to see the possible quantizations. We emphasize that $c_\phi$ and $c_F$ are parameters that interpolate different field theories living on the boundary. In general we can impose two independent mixed boundary conditions among $\delta \alpha, \delta \beta, \delta \mu, \delta Q_F$. The need for the general mixed boundary condition become more obvious in the massless case \S \ref{sec:MasslessVariationExponentialW} and the next section \S \ref{sec:BoundaryProblemPolynomialW}.   

Let us consider some particular examples for simplicity. When $c_\phi=0$ and $c_F=0$, it is required to fix $\alpha$ and $\mu$ at the boundary. The dual field theory operators have the expectation values  
\begin{align}
\langle \mathcal O_\alpha \rangle =  \frac{(\lambda_+ - \lambda_-)}{\tilde \kappa^2} \beta\;, \qquad 
 \langle \mathcal O_\mu \rangle =  - q \;.
\end{align}
For a different choice, $c_\phi=1$ and $c_F=1$, it is required to fix $\beta$ and $Q_F$ at the boundary. Then
\begin{align} \label{WExpFixingBeta}
\langle \mathcal O_\beta \rangle =  \frac{(\lambda_- - \lambda_+)}{\tilde \kappa^2} \alpha\;, \qquad 
 \langle \mathcal O_{Q_F} \rangle =  \frac{\mu q}{Q_F} \;.
\end{align}
For $c_F=0$ with fixed $\mu$, one can have the following quantization condition for the scalar 
\begin{align} \label{GeneralVariationAboveBFBoundExponentialW}
[c_\phi-1]  \beta \delta\alpha +  c_\phi  \alpha \delta\beta=0 \;.
\end{align} 
As we see below with a particular EMD background, a requirement from the thermodynamic first law fixes the parameter $c_\phi$ to be $1/3$.

Before moving on, let us consider $\alpha=0$ in \eqref{VarExpGen}. 
\begin{align} 
\delta S_M 
& \propto ~
\frac{(\Lambda_\phi - c_\phi \lambda_- - [c_\phi-1] \lambda_+ ) \beta \delta\alpha }{\tilde \kappa^2 }+ (c_F -1) \mu q \frac{\delta \mu}{\mu} + c_F \mu q \frac{\delta Q_F}{Q_F} = 0 \;.
\end{align} 
If we choose $\mu =const. $ for $c_F=0$ or $ Q_F =const.$ for $c_F=1$, the scalar expectation value is  
\begin{align}
\langle \mathcal O_{\alpha=0} \rangle =  \frac{\Lambda_\phi + [1-c_\phi]\lambda_+ - c_\phi \lambda_-}{\tilde \kappa^2} \beta  \;.
\end{align}
Thus the expectation value depends on both the boundary terms $\Lambda_\phi, c_\phi$ and is not fixed. This happens also for the scalar with the mass saturated at the BF bound as we see below.

\subsubsection{Scalar mass saturating the BF bound} \label{sec:ExponentialWBFBound}

For general $c_\phi$ and $c_F$, the variation \eqref{GeneralBoundaryTerms} has the following form for the scalar mass saturating the BF bound \eqref{ScalarFalloffBFBound} 
\begin{align} 
\begin{split}
\delta S_M 
&= \int d^d x \left\{ \frac{\{ [\Lambda_\phi -(2c_\phi -1) d/2] \log r + (2c_\phi -1)\} (\alpha \log r + \beta) -\beta (c_\phi -1)}{\tilde \kappa^2 }  \delta \alpha  \right. \\
&\qquad\qquad + \frac{[\Lambda_\phi -(2c_\phi -1) d/2]  (\alpha \log r + \beta) + (c_\phi -1) \alpha}{\tilde \kappa^2 } \delta \beta \\
&\qquad\qquad \left.+ (c_F -1) \mu q \frac{\delta \mu}{\mu} + c_F \mu q \frac{\delta Q_F}{Q_F}\right\} \;.
\end{split}
\end{align}
Again the term \eqref{NewTermExponentialCoupling} decays quickly compared to the other terms. In general we can impose two independent conditions among $\delta \alpha, \delta \beta, \delta \mu, \delta Q_F$. Yet, before that, we need to take care of the divergent parts. We find the following choice works best. 
\begin{align} \label{ConditionExponentialWBFBoundAlpha}
\Lambda_\phi = (2c_\phi -1) \left( \frac{d}{2}  -  \frac{1}{  \log r} \right) \;.
\end{align}
Then
\begin{align} 
- \frac{ (c_\phi -1) \beta \delta \alpha + c_\phi \alpha \delta \beta }{\tilde \kappa^2 }  +(c_F -1) \mu q \frac{\delta \mu}{\mu} + c_F \mu q \frac{\delta Q_F}{Q_F} = 0 \;.
\end{align} 
For this to satisfy, we can impose two mixed conditions in general. 

There are various ways to find specific cases for the well defined variational problem. 
It works, for example, if one imposes the mixed boundary condition 
\begin{align} 
 (c_\phi -1) \beta \delta \alpha + c_\phi \alpha \delta \beta =0 \;,
\end{align}
for the scalar contribution. This include the special case for $c_\phi=0$. For this case with $c_F=0$, it is required to fix $\alpha$ and $\mu$ at the boundary. The corresponding field theory dual operators have the following expectation values  
\begin{align}
\langle \mathcal O_\alpha \rangle =  \frac{\beta}{\tilde \kappa^2}\;, \qquad 
 \langle \mathcal O_\mu \rangle =  - q \;.
\end{align}

When $\alpha=0$, we require to set 
\begin{align} \label{ConditionExponentialWBFBoundAlphaZero}
\Lambda_\phi =(2c_\phi -1) d/2 
\end{align}
in \eqref{VarExpGen}, then the variation becomes 
\begin{align}  \frac{c_\phi \beta}{\tilde \kappa^2 }  \delta \alpha   + (c_F -1) \mu q \frac{\delta \mu}{\mu} + c_F \mu q \frac{\delta Q_F}{Q_F} = 0 \;.
\end{align}
Thus we impose $\alpha=const.$ and $\mu = const.$ for $c_F=0$ to find 
\begin{align} \label{ExpectationValuesExponeitialWBFBoundAlphaZero}
\langle \mathcal O_\alpha \rangle =  \frac{c_\phi \beta}{\tilde \kappa^2}\;, \qquad 
 \langle \mathcal O_\mu \rangle =  - q \;.
\end{align}
In this case, the terms $\sqrt{-\gamma} \phi^2$ and $ \sqrt{-\gamma} \phi n^r \partial_r \phi $ are finite at the boundary. One of the parameters $c_\phi$ and $\Lambda_\phi$ can not be fixed from the variational problem and the holographic renormalization. Different parameters are associated with different field theories at the boundary. This indicates that fixing the boundary terms are intrinsically ambiguous when the scalar mass saturates the BF bound. This is also reflected in an example below.

\subsubsection{Massless scalar }  \label{sec:MasslessVariationExponentialW}

For a massless scalar, we have $\lambda_- = 0$ and $ \lambda_+=d $. \eqref{GeneralBoundaryTerms} gives  for general $c_\phi$ 
\begin{align}  \label{MasslessVariationExponentialW}
\begin{split}
\delta S_M &= \int d^d x \left\{ \left[ \frac{ \Lambda_\phi \alpha}{\tilde \kappa^2 r^{-d}} +\frac{(\Lambda_\phi -(c_\phi -1) d) \beta}{\tilde \kappa^2}   + c_F \mu q c_W \right] \delta \alpha  \right. \\
&\qquad\qquad+  \left[ \frac{ (\Lambda_\phi -d c_\phi) \alpha}{\tilde \kappa^2 r^{-d}} +\frac{(\Lambda_\phi -(2c_\phi -1) d) \beta}{\tilde \kappa^2}   + c_F \mu q c_W \right] \frac{\delta \beta}{r^{d}} \\
&\qquad\qquad \left. + (c_F -1) \mu q \frac{\delta \mu}{\mu} + c_F \mu q \frac{\delta Q_F}{Q_F}\right\}\;.
\end{split}
\end{align}
The new term \eqref{NewTermExponentialCoupling} actually contributes and provides interesting options. 
 
Let us impose $\Lambda_\phi=0$. Then the expression simplifies to 
\begin{align}  \label{MasslessVariationExponentialWLambdaZero}
\left[\frac{(1-c_\phi ) d \beta}{\tilde \kappa^2}   + c_F \mu q c_W \right]  \delta \alpha -\frac{d c_\phi \alpha}{\tilde \kappa^2} \delta \beta  + (c_F -1) \mu q \frac{\delta \mu}{\mu} + c_F \mu q \frac{\delta Q_F}{Q_F} =0 \;.
\end{align} 
For the grand canonical case, $c_F=0$, it is natural to treat the variational problem between the gauge and scalar fields separately. The case $c_F \neq 0$ gives us new possibilities. From the condition \eqref{MasslessVariationExponentialWLambdaZero}, it is reasonable to fix two mixed conditions among $\delta \alpha, \delta \beta, \delta \mu, \delta Q_F$. For example,  if we choose $c_\phi=1$, it is natural to impose a mixed condition among $\delta \alpha$, $\delta \mu$ and $\delta Q_F$ as  
\begin{align}  
\mu q \left[ c_F c_W  \delta \alpha  + (c_F -1)  \frac{\delta \mu}{\mu} + c_F  \frac{\delta Q_F}{Q_F}\right]  =0 \;, \qquad \& \qquad \delta \beta=0  \;.
\end{align} 

Let us mention some particular cases that can be done by fixing the parameters. If one chooses $c_\phi=0$ and $c_F=1$, the variational problem is well defined for $\alpha=const.$ and $Q_F=const.$, and the corresponding expectation values are 
\begin{align}
\langle \mathcal O_{\alpha} \rangle = \frac{ d \beta}{\tilde \kappa^2}   + \langle \mathcal O_{Q_F} \rangle Q_F c_W  \;, \qquad 
 \langle \mathcal O_{Q_F} \rangle =  \mu q \frac{\delta Q_F}{Q_F} \;.
\end{align}
Thus the expectation value of the dual scalar operator is a function of the expectation value of the dual current operator. 
Another case, $c_\phi=1$ and $c_F =1$, provides a rather different situation in contrast to the Maxwell case, $W=1/4$. This happens due to the presence of the term $c_W \mu q $. One simple choice is to set $\alpha=0$ for the well defined variational problem. Then 
\begin{align} \label{ParticularCaseAlphaZero}
\langle \mathcal O_{\alpha=0} \rangle =  \langle \mathcal O_{Q_F} \rangle Q_F c_W  \;, \qquad 
 \langle \mathcal O_{Q_F} \rangle =  \frac{\mu q }{Q_F} \;.
\end{align}
Now we check this is a particular case we consider momentarily for $\alpha=0$.  

Let us consider $\alpha=0$.  
\begin{align}  \label{AlphaZeroVariation}
\left[\frac{(\Lambda_\phi -(c_\phi -1) d) \beta}{\tilde \kappa^2}   + c_F \mu q c_W \right] \delta \alpha   + (c_F -1) \mu q \frac{\delta \mu}{\mu} + c_F \mu q \frac{\delta Q_F}{Q_F} =0 \;.
\end{align} 
Thus the variational problem is well defined for $Q_F=const.$ if $c_F=1$. The corresponding expectation values are 
\begin{align} \label{AlphaZeroVEV}
\langle \mathcal O_{\alpha=0} \rangle = \frac{(\Lambda_\phi -(c_\phi -1) d) \beta}{\tilde \kappa^2}   + \langle \mathcal O_{Q_F} \rangle Q_F c_W  \;, \qquad 
 \langle \mathcal O_Q \rangle =   \frac{\mu q }{Q_F} \;.
\end{align}
Note that the expectation value of the operator dual to the scalar depends on the undetermined parameters $\Lambda_\phi $ and $c_\phi $. This is consistent with and yet more general than the result \eqref{ParticularCaseAlphaZero}, where we have been forced to set $\alpha=0$ after choosing $\Lambda_\phi=0$ and $c_\phi=1$. 
This happens for the massless scalar with the boundary falloff $\phi = \frac{\beta }{r^d}$. Once again the boundary terms $\sqrt{-\gamma} \phi^2$ and $ \sqrt{-\gamma} \phi n^r \partial_r \phi $ are finite at the boundary. The coefficients $c_\phi, \Lambda_\phi$ of the boundary terms are not fixed by the variational problem.

\subsection{Polynomial coupling: $W(\phi) \sim \phi^k  $}  \label{sec:BoundaryProblemPolynomialW}

The case with the polynomial coupling $W(\phi) \sim \phi^k $ is more subtle. The boundary term $\frac{\partial W(\phi)}{\partial \phi}$ has the form 
\begin{align} \label{NewTermPolynomialCoupling}
\frac{2}{\kappa^2} \int d^d x \sqrt{-\gamma}  \frac{k}{\phi_s} W ~n_r F^{r\nu} A_\nu (\delta \phi) \;.
\end{align}
Here we treat this term semi-classically, meaning that $\delta \phi$ has the full variational property as $\delta \phi  = \frac{\delta \alpha }{r^{\lambda_-}} +  \frac{\delta \beta }{r^{\lambda_+}} $, while the rest of the term are evaluated according to a given particular solution. For example, $\phi_s = \frac{\phi_0}{r^{\lambda_-}} $ or $\phi_s = \frac{\phi_0}{r^{\lambda_+}} $. One might try to understand the variational problem with full non-linear properties of coupled scalar and gauge fields at the boundary, which is beyond the scope of this paper.  

We consider the polynomial coupling $W(\phi) = \frac{1}{4} W_0 \phi^k$ and the gauge field 
\begin{align}
A_t = \mu + \frac{1}{1-\lambda_{Q}}\frac{Q_F}{r^{\lambda_Q-1}} 
\end{align}
at the boundary. This gives the same form for $F_{rt}$ as in \eqref{FieldStrengthBoundaryQ}. 
From the conserved quantity \eqref{ConservedMaxwellCharge}, the scalar field behaves as 
\begin{align}
\phi_s = c_W r^{-\frac{d-\lambda_Q-1}{k}}\;, \qquad (c_W)^k = 2\kappa^2 L^{d-1} \frac{q}{Q_F} \;.
\end{align}
Note that the radial dependence of the scalar should be either $r^{-\lambda_+}$ or $r^{-\lambda_-}$. The coefficient $(c_W)^k$ determines the way to split the charge $q$ between the gauge field and the coupling $W$. $W_0$ cancels out once we use the relation \eqref{ConservedMaxwellCharge}. 
Here we consider three different cases for the scalar mass as in \S \ref{sec:BoundaryProblemExponentialW}.

\subsubsection{Scalar mass above the BF bound} 

Let us first consider $\phi_s = c_W r^{-\lambda_-}$. The scalar solution is realized by its slower falloff. For general $\Lambda_\phi$ and $c_\phi$, the variation \eqref{GeneralBoundaryTerms} has the following form, after using \eqref{GeneralScalarAsymptotics} and \eqref{FieldStrengthBoundaryQ}, 
\begin{align}  \label{VarPolynomialGenScalarLambdaMinus}
\delta S_M 
&= \int d^d x \left\{ 
\frac{(\Lambda_\phi + [1-2c_\phi] \lambda_-) \alpha \delta \alpha}{\tilde \kappa^2 r^{-d+2\lambda_-}} \right. \nonumber \\
&\qquad\qquad+\frac{\{ (\Lambda_\phi - c_\phi \lambda_- - [c_\phi-1] \lambda_+ ) \beta + \frac{k c_F}{c_W} \mu q \tilde \kappa^2 \}  \delta\alpha + (\Lambda_\phi - c_\phi \lambda_+ - [c_\phi-1] \lambda_- ) \alpha \delta\beta}{\tilde \kappa^2 } \nonumber \\
& \left. \qquad\qquad+ (c_F -1) \mu q \frac{\delta \mu}{\mu} + c_F \mu q \frac{\delta Q_F}{Q_F}\right\} \;.
\end{align} 
Note that the term \eqref{NewTermPolynomialCoupling} provides a finite contribution.
Here we impose 
\begin{align} \label{ConditionForAboveBFBoundPolynomialW} 
\Lambda_\phi + [1-2c_\phi] \lambda_- = 0\;,
\end{align}
to have a well defined variational problem. This also yield the on-shell action to be finite. 
After that, one can impose two conditions among $\delta \alpha, \delta \beta, \delta \mu, \delta Q_F$ for given $c_\phi$ and $c_F$. 

Instead, we consider some special cases. If one choose $c_F=0$, then the term \eqref{NewTermPolynomialCoupling} vanishes, and it is natural to fix $\mu$ at the boundary. The scalar and vector variations separate. We can use the mixed condition on the scalar variation for general $c_\phi$. 
\begin{align}
(\lambda_- -\lambda_+ ) [c_\phi-1]  \beta \delta\alpha + (\lambda_- -\lambda_+ ) c_\phi  \alpha \delta\beta =0 \;.
\end{align}
If we choose a special case $c_\phi=0$, it is required to fix $\alpha$ at the boundary. Then
\begin{align}
\langle \mathcal O_\alpha \rangle =  \frac{(\lambda_+ - \lambda_-)}{\tilde \kappa^2} \beta  \;, \qquad 
 \langle \mathcal O_\mu \rangle =  - q \;.
\end{align}

In contrast, the case $c_F \neq 1$ is different. Consider $c_F=1$ for simplicity. Then \eqref{VarPolynomialGenScalarLambdaMinus} with the condition \eqref{ConditionForAboveBFBoundPolynomialW} gives  
\begin{align} 
\left[ (\lambda_- -\lambda_+ ) [c_\phi-1] \frac{ \beta}{\tilde \kappa^2} +  \frac{k }{c_W} \mu q  \right] \delta\alpha + (\lambda_- -\lambda_+ ) c_\phi  \frac{\alpha}{\tilde \kappa^2} \delta\beta 
+  \mu q \frac{\delta Q_F}{Q_F}=0 \;.
\end{align} 
Let us consider this in detail with some special cases. 
For $c_\phi=0$, it is natural to fix $\alpha$ and $Q_F$ at the boundary. Then 
\begin{align}
\langle \mathcal O_\alpha \rangle =  \frac{(\lambda_+ - \lambda_-)}{\tilde \kappa^2} \beta - \frac{k }{c_W} Q_F  \langle \mathcal O_{Q_F} \rangle  \;, \qquad 
 \langle \mathcal O_{Q_F} \rangle =  \frac{\mu  q }{Q_F} \;.
\end{align}
The expectation value $\langle \mathcal O_\alpha \rangle$ for $c_\phi=0$ depends not only $\beta$, but also $Q_F  \langle \mathcal O_{Q_F} \rangle $.

For $c_\phi=1$, fixing $\beta$ is no longer an option due to the presence of the term proportional to $\mu q$. Actually there is a more general mixed condition involving $\delta \alpha$ and $\delta Q_F$. The variation is 
\begin{align} \label{ScalarAboveBFCoditionedVariationCphiOne}
\frac{(\lambda_- -\lambda_+ )   \alpha \delta\beta}{\tilde \kappa^2 } + \mu q \left( \frac{k}{c_W} \delta\alpha +  \frac{\delta Q_F}{Q_F} \right) =0 \;.
\end{align}  
We need to impose two conditions. It is possible to choose $\beta=const.$ and a mixed condition 
\begin{align} 
 \frac{k}{c_W} \delta\alpha +  \delta \log Q_F  =0 \;.
\end{align} 
Thus $Q_F$ variation is tied with that of $\alpha$. Of course, we should not fix $\alpha$ because it is too restrictive. This signifies a new possibility to have a mixed boundary condition between the scalar and gauge variations.  \\

Now we come to the other scalar solution, $\phi_s = c_W r^{-\lambda_+}$. For general $\Lambda_\phi$ and $c_\phi$,  \eqref{GeneralBoundaryTerms} gives
\begin{align}  \label{VarPolynomialGenScalarLambdaPlus}
\delta S_M 
&= \int d^d x \left\{ 
\frac{\{ (\Lambda_\phi + [1-2c_\phi] \lambda_-) \alpha +  k c_F \mu q \tilde \kappa^2/c_W \} \delta \alpha}{\tilde \kappa^2 r^{-d+2\lambda_-}} \right. \nonumber \\
&\qquad\qquad+\frac{ (\Lambda_\phi - c_\phi \lambda_- - [c_\phi-1] \lambda_+ ) \beta  \delta\alpha + \{ (\Lambda_\phi - c_\phi \lambda_+ -  [c_\phi-1] \lambda_- ) \alpha + k c_F \mu q \tilde \kappa^2/c_W \} \delta\beta}{\tilde \kappa^2 } \nonumber \\
& \left. \qquad\qquad+ (c_F -1) \mu q \frac{\delta \mu}{\mu} + c_F \mu q \frac{\delta Q_F}{Q_F}\right\} \;.
\end{align} 
To make things a little more clear, let us fix $c_F=1$ and $Q_F =const.$ first. Then the expectation value for the operator dual to $Q_F$ is $\langle \mathcal O_{Q_F} \rangle =  \frac{\mu q}{Q_F}$. 
We consider the case $\langle \mathcal O_{Q_F} \rangle=const.$ and 
\begin{align} \label{ConditionForAboveBFBoundPolynomialLambdaPlus} 
\alpha = -  \frac{k \tilde \kappa^2}{c_W} \frac{Q_F \langle \mathcal O_{Q_F} \rangle }{\Lambda_\phi  + [1-2c_\phi] \lambda_-} \;, 
\end{align}
which is constant. The on-shell action is finite. We further require $c_\phi =0 $ to have a well defined variational problem. The resulting  expectation value is given by 
\begin{align}
\langle \mathcal O_{\alpha} \rangle =  \frac{(\Lambda_\phi + \lambda_+ )}{\tilde \kappa^2} \beta  \;.
\end{align}
If $c_\phi \neq 0$, the variational problem is too restrictive for $c_F=1$. Of course, we can choose more general mixed boundary condition including $c_F \neq 1$.

\subsubsection{Scalar mass saturating the BF bound} \label{sec:PolynomialWBFBound}

Here we again consider two separate cases depending on the scalar solution, either $\phi_s
= c_W r^{-d/2}$ or $\phi_s = c_W r^{-d/2} \log r$. Let us focus on $\phi_s = c_W r^{-d/2}$. For general $c_\phi$ and $c_F$, the variation \eqref{GeneralBoundaryTerms} gives, after using  \eqref{ScalarFalloffBFBound} 
\begin{align} \label{VariationPolynomialWBFBound}
\delta S_M 
&=\int d^d x \left\{  \frac{( [ \Lambda_\phi+ (1-2c_\phi)d/2] \log r + 2c_\phi -1 ) (\alpha \log r + \beta) -(c_\phi-1)\beta +  c_F \tilde \kappa^2 \mu q k \log r/c_W}{\tilde \kappa^2 }  \delta \alpha \right. \nonumber  \\
&\qquad\qquad +\frac{ ( \Lambda_\phi+[1-2c_\phi]d/2)(\alpha \log r + \beta) +(c_\phi-1)\alpha + c_F \tilde \kappa^2 \mu q k/c_W}{\tilde \kappa^2 }  \delta \beta  \\
&\qquad\qquad \left.+ (c_F -1) \mu q \frac{\delta \mu}{\mu} + c_F \mu q \frac{\delta Q_F}{Q_F}\right\} \;.\nonumber 
\end{align}
If one chooses $c_F=0$, the variational problem reduces to the case we considered before in \S \ref{sec:ExponentialWBFBound}. 

Let us focus on the case with $c_F \neq 0$, specifically $c_F=1$. We impose the condition 
\begin{align}
 \Lambda_\phi = (2c_\phi -1)d/2   \;.
\end{align} 
Then \eqref{VariationPolynomialWBFBound} gives 
\begin{align} 
 \frac{c_\phi \beta + \{ ( 2c_\phi -1 ) \alpha  + \tilde \kappa^2 \mu q k /c_W \} \log r}{\tilde \kappa^2 }  \delta \alpha  +\frac{ (c_\phi-1)\alpha +  \tilde \kappa^2 \mu q k/c_W}{\tilde \kappa^2 }  \delta \beta  + \mu q \frac{\delta Q_F}{Q_F} = 0 \;.
\end{align} 
In general, we need to impose two boundary conditions among $\delta \alpha, \delta \beta$ and $\delta Q_F $. 

Once we consider special cases, we see that only $c_\phi =0$ or $c_\phi =1$ can work for $c_F=1$ due to the non-trivial coefficients of $\delta \alpha$ and $\delta \beta$. 
If one chooses $c_\phi =0$, the variation is proportional to $ \delta \alpha \log r + \delta \beta$. Similar case was considered in \cite{Marolf:2006nd}. 
\begin{align} 
\frac{\{\tilde \kappa^2 \mu q k /c_W  - \alpha\} (\delta \alpha \log r+\delta \beta)}{\tilde \kappa^2 }   + \mu q \frac{\delta Q_F}{Q_F}= 0 \;.
\end{align} 
We can fix $\alpha =  \frac{  \tilde \kappa^2 k }{ c_W} \mu q = const.$ and impose the condition $\delta Q_F  = 0$. 
If we consider $c_\phi=1$. This brings the variation to the form
\begin{align} 
\frac{\beta + \{  \alpha  + \tilde \kappa^2 \mu q k /c_W \} \log r}{\tilde \kappa^2 }  \delta \alpha + \mu q \frac{  k}{c_W}  \delta \beta  + \mu q \frac{\delta Q_F}{Q_F}=0 \;.
\end{align} 
Now we can fix $\alpha = - \frac{  \tilde \kappa^2 k }{ c_W} \mu q = const.$ and impose the variational condition 
\begin{align}
\delta \alpha = 0 \;, \qquad 
\delta \log Q_F  = -\frac{k}{c_W} \delta \beta   \;.
\end{align}
Thus the variation of $Q_F$ is directly related to that of $\beta$, which is not fixed at the boundary. The expectation value of the scalar is also given by $\beta$.  
\begin{align}
\langle \mathcal O_{\alpha} \rangle = \frac{\beta}{\tilde \kappa^2  }  \;.
\end{align}

Before moving on, it is interesting to examine the case $\alpha=0$. 
The variation \eqref{VariationPolynomialWBFBound} gives 
\begin{align} 
\delta S_M 
&=\int d^d x \left\{  \frac{ \{ ( \Lambda_\phi+[1-2c_\phi]d/2) \beta + c_F \tilde \kappa^2 \mu q k/c_W \}  \log r -c_\phi \beta}{\tilde \kappa^2 }  \delta \alpha   \right. \nonumber  \\
&\qquad\qquad \left.  +\frac{ ( \Lambda_\phi+[1-2c_\phi]d/2) \beta + c_F \tilde \kappa^2 \mu q k/c_W}{\tilde \kappa^2 }  \delta \beta + (c_F -1) \mu q \frac{\delta \mu}{\mu} + c_F \mu q \frac{\delta Q_F}{Q_F}\right\} \nonumber\\
&=\int d^d x \left\{  \frac{ -c_\phi \beta}{\tilde \kappa^2 }  \delta \alpha   + (c_F -1) \mu q \frac{\delta \mu}{\mu} + c_F \mu q \frac{\delta Q_F}{Q_F}\right\} \;.
\end{align}
In the last line, we use the relation 
\begin{align} \label{ConditionMixedBCPolynomialWBFBound}
( \Lambda_\phi+[1-2c_\phi]d/2) \beta + \frac{c_F}{c_W} k \tilde \kappa^2 \mu q  = 0\;, 
\end{align}
which is to be understood that $\mu q$ are adjusted to satisfy the relation for general $\beta$. 
The variational problem fixes $\alpha=0$ and requires the mixed condition 
\begin{align} \label{MixedBCPolynomialWBFBound}
\delta \alpha = 0 \;, \qquad   (c_F -1) \delta \log \mu + c_F  \delta \log Q_F =0 \;.
\end{align}
The vacuum expectation value of the operator dual to the scalar is 
\begin{align}
\langle \mathcal O_{\alpha=0} \rangle = -\frac{ c_\phi }{\tilde \kappa^2 } \beta \;.
\end{align} 
This case is actually realized in an example below with the condition \eqref{ConditionMixedBCPolynomialWBFBound} satisfied. \\

Let us be brief on the other case $\phi_s = c_W r^{-d/2} \log r$. Again, the variation \eqref{GeneralBoundaryTerms} has the form for general $c_\phi$ and $c_F$   
\begin{align} 
\delta S_M 
&=\int d^d x \left\{  \frac{( [(1-2c_\phi)d/2 + \Lambda_\phi] \log r + 2c_\phi -1 ) (\alpha \log r + \beta) -(c_\phi-1)\beta + c_F \tilde \kappa^2 \mu q k /c_W}{\tilde \kappa^2 } \delta \alpha   \right. \nonumber  \\
&\qquad\qquad + \frac{ ([1-2c_\phi]d/2 + \Lambda_\phi)(\alpha \log r + \beta) +(c_\phi-1)\alpha }{\tilde \kappa^2 } \delta \beta   \\
&\qquad\qquad \left.+ (c_F -1) \mu q \frac{\delta \mu}{\mu} + c_F \mu q \frac{\delta Q_F}{Q_F}\right\} \;.\nonumber 
\end{align}
For $c_F=0$, one can refer to \S \ref{sec:ExponentialWBFBound}. $c_F \neq 0$ provides various possibilities.  
For $c_F=1$, we set $\Lambda_\phi = 0$ and $c_\phi =1/2$. Then 
\begin{align}  
\ \frac{\beta/2 +  \tilde \kappa^2 \mu q k /c_W}{\tilde \kappa^2 }  \delta \alpha  - \frac{ \alpha/2 }{\tilde \kappa^2 } \delta \beta   + \mu q \frac{\delta Q_F}{Q_F}=0\;. 
\end{align} 
In general, we can impose two conditions. For example $\alpha=const.$ and one mixed condition $\frac{ \alpha/2 }{\tilde \kappa^2 } \delta \beta  = \mu q \frac{\delta Q_F}{Q_F} $ by coupling the variation of $Q_F$ to the variation of $\delta \beta$. 
Another simple case is to fix $\beta  = - \frac{2 k }{c_W} \langle \mathcal O_{Q_F} \rangle Q_F $ with condition $ \langle \mathcal O_{Q_F} \rangle =const.$, then 
\begin{align} 
\langle \mathcal O_{\beta} \rangle = -\frac{\alpha }{ 2 \tilde \kappa^2 }  \;, \qquad 
 \langle \mathcal O_{Q_F} \rangle =   \frac{\mu  q }{Q_F} \;.
\end{align}

\subsubsection{Massless scalar}

For the massless scalar, there are two solutions $\phi_s \sim r^0, r^{-d} $. 
Let us start with $\phi_s=c_W$.  Then 
\begin{align} \label{VariationMasslessPolynomial333}
\begin{split}
\delta S_M &= \int d^d x \left\{ \frac{ \Lambda_\phi \alpha \delta \alpha}{\tilde \kappa^2 r^{-d}} +\frac{[(\Lambda_\phi - (c_\phi -1)d) \beta + c_F \mu q k \tilde \kappa^2/c_W ]\delta \alpha   + (\Lambda_\phi - c_\phi d) \alpha \delta  \beta}{\tilde \kappa^2} \right. \\
&\qquad\qquad \left.+ (c_F -1) \mu q \frac{\delta \mu}{\mu} + c_F \mu q \frac{\delta Q_F}{Q_F}\right\}\;.
\end{split}
\end{align}  
We can impose two general mixed condition. For simplicity, we consider $c_F=1$ below. 

For $\Lambda_\phi=0$, the variation \eqref{VariationMasslessPolynomial333} simplifies to 
\begin{align} \label{EMDZeroMassPolynomialW}
\frac{[(1-c_\phi )d \beta + \mu q k \tilde \kappa^2/c_W ]\delta \alpha  - c_\phi d \alpha \delta  \beta}{\tilde \kappa^2} + \mu q \frac{\delta Q_F}{Q_F} = 0\;.
\end{align}
One can work out a variation for general $c_\phi$. We consider some specific cases.    
For $c_\phi=0$, a simple choice is to fix $\alpha= const.$ and $Q_F = const.$. Then
\begin{align}
\langle \mathcal O_\alpha \rangle =  \frac{d }{\tilde \kappa^2} \beta    + \langle \mathcal O_{Q_F} \rangle Q_F  \frac{k}{ c_W} \;, \qquad 
 \langle \mathcal O_{Q_F} \rangle =   \frac{\mu q }{Q{Q_F} } \;.
\end{align}
It is interesting to see that the scalar expectation value depends on that of the charge operator due to the coupling term $WF^2$. 
Similarly, for $c_\phi=1$, we have 
\begin{align} \label{PartiaularExampleAlphaZeroPolyWMassless1}
\frac{- d \alpha \delta  \beta}{\tilde \kappa^2} + \mu q \frac{  k  \delta \alpha  }{c_W} +  \mu q \frac{\delta Q_F}{Q_F} = 0\;.
\end{align}  
One can impose a mixed condition as familiar from previous examples
\begin{align}
\delta \beta = 0 \;, \qquad  k  \delta \alpha + c_W \delta \log Q_F  = 0 \;.
\end{align}

Finally, we consider the case $\phi_s=c_W r^{-d}$.  
\begin{align} \label{VariationPolynomialWMassless333}
\begin{split}
\delta S_M &= \int d^d x \left\{ \frac{ [\Lambda_\phi \alpha + c_F \mu q k \tilde \kappa^2/c_W ]\delta \alpha}{\tilde \kappa^2 r^{-d}} \right. \\
&\qquad\qquad \left. +\frac{(\Lambda_\phi - (c_\phi -1)d) \beta \delta \alpha   + [(\Lambda_\phi - c_\phi d) \alpha + c_F \mu q k \tilde \kappa^2/c_W ] \delta  \beta}{\tilde \kappa^2} \right. \\
&\qquad\qquad \left.+ (c_F -1) \mu q \frac{\delta \mu}{\mu} + c_F \mu q \frac{\delta Q_F}{Q_F}\right\}\;.
\end{split}
\end{align}  
For $c_F=1$, we fix $Q_F =const.$ and consider the expectation value for the operator dual to $Q_F$ is constant 
\begin{align}
\langle \mathcal O_{Q_F} \rangle =  \frac{\mu  q }{Q_F} = const.\;.
\end{align}
We impose 
\begin{align}
\alpha =- \frac{k \tilde \kappa^2}{\Lambda_\phi c_W} Q_F \langle \mathcal O_{Q_F} \rangle \;.
\end{align}
Then, the expression \eqref{VariationPolynomialWMassless333} simplifies to 
\begin{align} 
\frac{(\Lambda_\phi - (c_\phi -1)d) \beta \delta \alpha   - c_\phi d \alpha \delta  \beta}{\tilde \kappa^2} + \mu q \frac{\delta Q_F}{Q_F}  =0 \;.
\end{align}  
Let's consider some specific examples. 
For $c_\phi=0$, it is straightforward to fix $\alpha= const.$ and $Q_F = const.$ to find
\begin{align}
\langle \mathcal O_{\alpha} \rangle =  \frac{\Lambda_\phi + d }{\tilde \kappa^2} \beta   \;, \qquad 
 \langle \mathcal O_{Q_F} \rangle =   \frac{\mu q }{Q_F} \;.
\end{align}
If we choose $\Lambda_\phi = (c_\phi-1)d$, we can fix $\beta= const.$ and $Q_F = const.$ to find
\begin{align}
\langle \mathcal O_{\beta} \rangle =  \frac{-c_\phi d }{\tilde \kappa^2} \alpha   \;, \qquad 
 \langle \mathcal O_{Q_F} \rangle =   \frac{\mu q }{Q_F} \;.
\end{align}
Note that both the cases the expectation values of the scalar are functions of the unfixed parameters.

\section{EMD Solutions} \label{sec:EMDSolutions}

In this section, we apply the general programs of the boundary variational problem, the on-shell action and the stress energy tensors to some analytic EMD solutions with asymptotic AdS boundary. It is pleasant to check that all these programs fit together nicely.

\subsection{AdS$_4$ Background}  \label{sec:EMDAdS4}

The AdS$_4$ solution considered in \cite{Gubser:2009qt} has the following action and metric 
\begin{align} \label{AdS4EMDAction}
\begin{split}
S &=\frac{1}{2\kappa^2} \int d^4 x \sqrt{-g} \left[\mathcal R - W(\phi) F^2 - \frac{1}{2} (\partial \phi)^2 - V(\phi)  \right] \;, \\
ds^2 &= e^{2C} (-h dt^2 + d\vec{x}^2) + \frac{e^{-2C}}{h} dr^2 \;, \\ 
W &(\phi) = \frac{1}{4} e^{\phi/\sqrt{3}} \;, \quad 
V(\phi)= - \frac{6}{L^2} \cosh (\phi/\sqrt{3}) \;, \\
C &= \log \left(\frac{r }{L}\right) + \frac{3 }{4} \log\left( 1 + \frac{Q}{r} \right)\;,  \qquad  
h = 1 - \frac{\omega L^2}{ (Q+r)^3}\;, \\
F &= dA \;, \quad A = 
\left(-\frac{\sqrt{3 Q \omega}}{ Q+r} + \frac{\sqrt{3 Q} ~\omega^{1/6} }{ L^{2/ 3}}\right) dt \;, \quad
\phi = \frac{\sqrt{3}}{ 2} \log\left(1+ \frac{Q}{r}\right) \;. 
\end{split}
\end{align}
One can check the equations of motion \eqref{EMDEOM} are satisfied. 
As noted in \cite{Gubser:2009qt}, this solution has a naked singularity in the extremal limit $ \omega =Q^3/L^2$. Note $Q \neq Q_F$. There have been extensive literature works that have dealt with this issue 
\cite{Gubser:2000nd}-\cite{Gursoy:2008za},\cite{Charmousis:2010zz}. 
This solution can be uplifted to resolve the singularity by including stringy degrees of freedom as noted in \cite{Gubser:2009qt}. A solution with similar potential in asymptotic AdS is also analyzed in \cite{Hartnoll:2011pp}.

From the gauge field $A$, we can see 
\begin{align}
A_t 
= 
\frac{ \sqrt{3Q} \omega ^{1/6}}{L^{2/3}} - \frac{\sqrt{3Q \omega }}{r} + \mathcal O(r^{-2}) \;. 
\end{align}
The constant term is the chemical potential $\mu = \frac{ \sqrt{3Q} \omega ^{1/6}}{L^{2/3}} $, and the coefficient of the second term is proportional to the charge density, $q = \frac{1}{2 \kappa ^2}\frac{\sqrt{3 Q \omega }}{L^2}$. Thus $\mu q = \frac{3Q}{2 \kappa ^2}\frac{\omega^{2/3}}{L^{8/3}} $. Depending on the choice of the boundary term, we can either fix the chemical potential or the charge. 

The mass of the scalar can be evaluated  
\begin{align} \label{MassAdS4Gubser}
m^2_\phi L^2 = - \frac{\partial^2}{\partial \phi^2} \left(  6 \cosh \left(\frac{\phi}{\sqrt{3}}\right)- \frac{L^2}{4} e^{\frac{\phi}{\sqrt{3}}} F^2 \right)\bigg|_{\phi=0} = -2  \;.
\end{align}
The gauge field term decay sufficiently fast and does not contribute to the mass. 
Thus the scalar field has $\lambda_-=1$ and $\lambda_+=2$ in \eqref{GeneralScalarAsymptotics}. They are slightly above the BF bound. Both of the scalar falloffs are normalizable. Note that the particular solution is supported by the slower falloff of the scalar field 
\begin{align} \label{AdS4ScalarFalloff}
\phi = \frac{\sqrt{3}}{ 2} \frac{Q}{r} + \mathcal O(r^{-2}) \;.
\end{align}

The temperature and entropy density can be readily evaluated   
\begin{align}
T = \frac{3 \mu  \sqrt{-Q+\left(L^2 \omega \right)^{1/3}}}{4 \pi  \left(L^2 \omega \right)^{5/6}} \;, \qquad 
s = \frac{2 \pi \mu ^{1/2} \sqrt{ -Q+\left(L^2 \omega \right)^{1/3}}}{L \kappa ^2} \;.
\end{align}
We express physical quantities in terms of $Q, \omega$ using $(Q+ r_h)^3 = \omega L^2 $ where $r_h$ is a horizon radius. 

\subsubsection{On-shell action} 

Let us evaluate the on-shell action following \S \ref{sec:OnShellAction}. From the Einstein equation, we get  
\begin{align}
R= \frac{1}{2} (\partial \phi )^2 - \frac{12}{L^2} \cosh (\phi/\sqrt{3})\;,
\end{align}
which can be used to evaluate to find   
\begin{align}
S_{on-shell} &=\frac{1}{2\kappa^2} \int d^4 x \int_{r_h}^{r_\epsilon} dr\sqrt{-g}  ~\frac{2}{3}\left[V-W F^2 \right] + S_{b} \;,  
\end{align}
where the boundary terms include the Gibbons-Hawking term, the Balasubramanian-Kraus terms, and the scalar and vector boundary terms 
\begin{align}
S_{b} &= - \frac{1}{\kappa^2} \int_{ \partial M} d^4 x \sqrt{-\gamma} 
\left[ \Theta + \frac{2}{L} - \frac{L}{2} \mathcal R^{(3)} -\frac{c_\phi}{2} \phi n^r \partial_r \phi -\frac{\Lambda_\phi}{4 L} \phi^2 
- 2 c_F W n_r F^{rt} A_t \right] \;.
\end{align}

To have a finite on-shell action, we impose the condition 
\begin{align}  \label{ConditionForAboveBFBoundExponentialW22} 
\Lambda_\phi = 2 c_\phi -1 \;.
\end{align}
This is consistent with the condition \eqref{ConditionForAboveBFBoundExponentialW}, with $\lambda_-=1$, that is used to make the variational problem well defined. 
Then we get 
\begin{align} \label{OnShellActionExponentialW} 
s_{on-shell} &=  \frac{(1-3c_\phi) Q^3 - 48 c_F Q \omega^{2/3} L^{4/3} +16 L^2 \omega }{32 L^4 \kappa ^2}\;.
\end{align}
Here $s_{on-shell}$ is a density, the on-shell action divided by the volume of the field theory coordinates including the compactified time.  
From this on-shell action (density), we identify the thermodynamic potential (density) 
\begin{align}
G = - \frac{(1-3c_\phi) Q^3 - 48 c_F Q \omega^{2/3} L^{4/3} +16 L^2 \omega }{32 L^4 \kappa ^2} \;.
\end{align}
Below we also check that this grand potential is identical to the pressure of the system. 

The corresponding stress energy tensor \eqref{FullStressEnergyTensor} is 
\begin{align} 
\kappa^2 T_{ij} &= \Theta_{ij} - \Theta \gamma_{ij} - \frac{2}{L} \gamma_{ij}
- L G^{(3)}_{ij} +\gamma_{ij} \left[\frac{c_\phi}{2} \phi n^r \partial_r \phi+ \frac{\Lambda_\phi}{4 L} \phi^2 \right] \\
&\qquad\qquad+ c_F \Big[ 2 \gamma_{ij} W n_r F^{rt} A_t -4 W n^r F_{r i} A_{j}  \Big] \;. \nonumber 
\end{align}
Explicit computation gives the following data (after imposing the condition $\Lambda_\phi = 2 c_\phi -1 $ given in \eqref{ConditionForAboveBFBoundExponentialW22} so that the stress energy tensor is finite) 
\begin{align} \label{StressEnergyTensorExponentialW}
\begin{split}
&E = \langle T_{tt} \rangle = \frac{(3c_\phi-1) Q^3 - 48 c_F Q \omega^{2/3} L^{4/3}+ 32 L^2 \omega }{32 \kappa^2 L^4 } \;,  \\ 
&P = \langle T_{xx} \rangle = \langle T_{yy} \rangle=  \frac{(1-3c_\phi) Q^3 - 48 c_F Q \omega^{2/3} L^{4/3} +16 L^2 \omega }{32 \kappa^2 L^4} \;.  
\end{split}
\end{align}
Here $E$ and $P$ are identified as energy density and pressure. The pressure is nothing but the grand potential, $P=-G$. 
 
One can also explicitly compute the mass density 
\begin{align} \label{MassExponentialW}
M &= \frac{(3c_\phi-1) Q^3 - 48 c_F Q \omega^{2/3} L^{4/3} + 32 L^2 \omega }{32 \kappa^2 L^4 } \;.  
\end{align}
Thus we confirm that the mass and the grand potential both depends on the parameters $c_\phi$ and $c_F$. 
At this point, one can readily check that the following thermodynamic relation holds if we set $c_F=0$.  
\begin{align} \label{ExactFormPotential}
\Omega = M - Ts - \mu q \;, \qquad \text{for} \quad c_F=0 \;.
\end{align}

\subsubsection{Grand canonical ensemble $c_F=0$} \label{sec:GrandcanonicalExponentialW}

For the grand canonical ensemble ($c_F=0$), we examine possible quantizations with the results in \S \ref{sec:ExponentialWAboveBFBound}. {\it A priori,} as far as the condition \eqref{ConditionForAboveBFBoundExponentialW22} is satisfied, all the possible quantizations are legitimate. The solution \eqref{AdS4EMDAction} is realized with the scalar field \eqref{AdS4ScalarFalloff}. We are going to see how the parameters $c_\phi$ is fixed for this solution. 

Let us start tentatively by examine the case $c_\phi = 0$. Then we get the following data for the dual field theory 
\begin{align}
\langle \mathcal O_{\alpha\sim Q=fixed} \rangle =  \frac{(\lambda_+ - \lambda_-)}{\tilde \kappa^2}  \beta=  \frac{1}{\tilde \kappa^2} \beta = 0\;, \qquad 
 \langle \mathcal O_\mu \rangle = -q  = -\frac{\sqrt{3 Q \omega }}{2 \kappa ^2 L^2} \;.
\end{align}
Where we use $\alpha =\frac{\sqrt{3}}{2}Q, \beta=0, \lambda_+=2, \lambda_-=1 $. Note $Q\neq Q_F$.
The corresponding on-shell action and thus the grand potential $\Omega$ are 
\begin{align} \label{GrandPotentialCphiZero}
\Omega = - s_{on-shell} &= - \frac{Q^3  +16 L^2 \omega }{32 L^4 \kappa ^2}\;, 
\end{align}
and the field theory stress energy tensors are 
\begin{align}
E=M = \langle T_{tt} \rangle &= \frac{-Q^3 + 32 L^2 \omega }{32 \kappa^2 L^4 } \;,  \qquad 
P = \langle T_{xx} \rangle = \langle T_{yy} \rangle=  \frac{ Q^3 + 16 L^2 \omega }{32 \kappa^2 L^4} \;.  
\end{align}
Here $E$ and $P$ are identified as energy and pressure. 
We check that the grand potential is nothing but the negative of the pressure, $\Omega=-P$. One can easily check the relation, $\Omega = M - Ts + \mu q$. In fact, it holds for any $c_\phi$ as in \eqref{ExactFormPotential}. 

Now there is a troublesome fact. It turns out that the grand potential \eqref{GrandPotentialCphiZero} does not satisfy the differential from of the first law   
\begin{align}
d\Omega = -s dT - q d\mu \;. 
\end{align}
If one think a little more, the reason is obvious. The term $Q^3$ is nowhere found in temperature, entropy, charge and chemical potential. Thus, even though this is legitimate quantization from the point of view of the variational problem, it is not acceptable from the point of view of the thermodynamic first law.  

Let us consider the general form of the grand potential with $c_\phi \neq 0$
\begin{align}
G = - \frac{(1-3c_\phi) Q^3 +16 L^2 \omega }{32 L^4 \kappa ^2} \;.
\end{align}
Now we can actually fix this parameter $c_\phi$ from the thermodynamic data. If one uses the differential form of the first law,  $d\Omega = -s dT - q d\mu $, 
then the parameter $c_\phi$ is uniquely fixed as 
\begin{align}
c_\phi = \frac{1}{3} \;, \qquad \Lambda_\phi = -\frac{1}{3} \;,
\end{align}
where $\Lambda_\phi$ is fixed by \eqref{ConditionForAboveBFBoundExponentialW22}. 
Thus it is required to put a mixed boundary condition discussed in   \eqref{GeneralVariationAboveBFBoundExponentialW}. 
Thus the differential form of the first law actually put more stringent constraint than that of the exact form. This demonstrates that the importance of the general boundary value problem along with the on-shell action to get consistent physical quantities.   

Summarizing, the AdS$_4$ theory is described by the grand potential $\Omega$ 
\begin{align}
\Omega  &= - \frac{\omega }{2 L^2 \kappa ^2}\;, 
\end{align}
and the field theory energy and pressure 
\begin{align}
E =  \frac{\omega }{\kappa^2 L^2 } \;,  \qquad 
P =  \frac{\omega }{2 \kappa^2 L^2} \;.  
\end{align} 
By demanding the differential form of the first law, all the parameters are fixed. 
One can also check that the trace condition is satisfied. 
\begin{align}
\langle T^\mu_{\ \mu} \rangle = -E + 2P = 0 \;.
\end{align}

\subsubsection{Semi canonical ensemble $c_F=1$} \label{sec:canonicalExponentialW}

The necessary information for the semi canonical ensemble is the same as that given in \S \ref{sec:GrandcanonicalExponentialW}. 
The on-shell action, from \eqref{OnShellActionExponentialW}, goes as 
\begin{align}
s_{on-shell} &=  \frac{(1-3c_\phi) Q^3 - 48 Q \omega^{2/3} L^{4/3} +16 L^2 \omega }{32 L^4 \kappa ^2}\;,
\end{align}
and the free energy 
\begin{align}
F = - \frac{(1-3c_\phi) Q^3 - 48 Q \omega^{2/3} L^{4/3} +16 L^2 \omega }{32 L^4 \kappa ^2} \;.
\end{align}
Thus it is consistent with the picture that the Helmholtz free energy is a Legendre transformation from the grand potential, 
\begin{align}
F = \Omega + \mu q = M - T s\;.
\end{align}
We further confirm that the differential form of the first law 
\begin{align}
dF = -s dT + \mu d q 
\end{align}
satisfies, again, for
\begin{align}
c_\phi = \frac{1}{3} \;.
\end{align}
Thus the semi canonical ensemble is well defined as a Legendre transform from the grand potential. 

For the rest of this sub-section, we comments on some results of the holographic renormalization of the semi canonical ensemble. 
The stress energy tensor can be obtained from \eqref{StressEnergyTensorExponentialW} by using $c_\phi =1/3, \Lambda_\phi =-1/3$. It has a further contribution from the boundary term compared to the grand canonical ensemble  
\begin{align}
\begin{split}
&E_F = \langle T_{tt} \rangle
= \frac{ 2 L^2 \omega -3 Q \omega^{2/3} L^{4/3} }{2 \kappa^2 L^4 } \;,  \\ 
&P_F = \langle T_{xx} \rangle = \langle T_{yy} \rangle 
=  \frac{  L^2 \omega - 3 Q \omega^{2/3} L^{4/3}}{2 \kappa^2 L^4} \;.  
\end{split}
\end{align}
Here $E_F$ and $P_F$ are identified as energy and pressure for semi canonical ensemble evaluated from the stress energy tensor. We note the pressure is again the negative of the free energy $P_F = -F$.  One can also explicitly compute the mass, \eqref{MassExponentialW}, to find 
\begin{align}
M_F &= E_F 
= \frac{ 2 L^2 \omega -3 Q \omega^{2/3} L^{4/3}}{2 \kappa^2 L^4 } \;.  
\end{align}
Now it is curious to find that this mass $M_F$ does not play the role of mass in the Helmholtz free energy. It will be interesting to figure out the meaning of the mass $M_F$ which is directly computed through the holographic renormalization for the semi canonical ensemble. 
For RNAdS black holes with a fixed charge, the energy is identified as the energy above the ground state (the extremal black hole) \cite{Chamblin:1999tk}\cite{Chamblin:1999hg}. Here we find that the energy of the fixed charge differs by the $\mu q$ not by the energy of the extremal black hole, $E_F = E - \mu q$.

\subsection{AdS$_5$ Background} \label{sec:EMDAdS5}

The AdS$_5$ solution considered in \cite{Gubser:2009qt} has the following action and metric 
\begin{align} \label{AdS5EMDAction}
\begin{split}
S &=\frac{1}{2\kappa^2} \int d^4 x \sqrt{-g} \left[\mathcal R - W(\phi) F^2 - \frac{1}{2} (\partial \phi)^2 - V(\phi)  \right] \;, \\
ds^2 &= e^{2C} (-h dt^2 + d\vec{x}^2) + \frac{e^{2D}}{h} dr^2 \;, \\
C &= \log \left(\frac{r }{L}\right) + \frac{1}{3 } \log\left( 1 + \frac{Q^2}{r^2} \right)\;,  \quad  
D= -\log \left(\frac{r }{L}\right) - \frac{2}{3 } \log\left( 1 + \frac{Q^2}{r^2} \right)\;,  \\
h &= 1 - \frac{\omega L^2}{ (Q^2+r^2)^2}\;, \quad W(\phi) = \frac{1}{4} e^{2\phi/\sqrt{6}} \;, \quad 
V(\phi)= -\frac{1}{L^2} \left(8 e^{\phi/\sqrt{6}} +4e^{-2\phi/\sqrt{6}}\right) \;, \\
F &= dA \;, \quad A = 
\left(-\frac{Q\sqrt{2\omega}}{ Q^2+r^2} + \frac{Q\sqrt{2 \omega}}{ Q^2+r_h^2}\right) dt \;, \quad
\phi = \frac{2}{ \sqrt{6}} \log\left(1+ \frac{Q^2}{r^2}\right) \;. 
\end{split}
\end{align}
The equations of motion for the metric, gauge and scalar fields, \eqref{EMDEOM}, are satisfied. 
The extremal limit is given by $ \omega =Q^3/L^2$. Note $Q\neq Q_F$. Again the system has a naked singularity at the extremal limit. Further discussions can be found in \S \ref{sec:EMDAdS4}. 

From the gauge field $A$, we have 
\begin{align}
A_t 
= \frac{ Q \sqrt{2}}{L} - \frac{ Q \sqrt{2\omega }}{r^2} + \mathcal O(r^{-2}) \;. 
\end{align}
The constant term is the chemical potential $\mu = \frac{ Q \sqrt{2}}{L} $ and the coefficient of the second term is proportional to the charge density $q = \frac{1}{\kappa ^2}\frac{Q\sqrt{2 \omega }}{L^3}$. The mass of the scalar comes from the potential term as $m^2_\phi L^2= -4 $. Thus the scalar field has the boundary behavior \eqref{ScalarFalloffBFBound} with $d=4$. It saturates the so-called BF bound. This is consistent with the radial fall-off of the scalar field 
\begin{align} \label{AdS5ScalarFalloff}
\phi = \sqrt{\frac{2}{3}} \frac{Q^2}{r^2} + \mathcal O(r^{-2}) \;.
\end{align} 
Compared to \eqref{ScalarFalloffBFBound}, this scalar solution realizes with the faster falloff at the boundary.

The temperature and entropy density are 
\begin{align}
T = \frac{\sqrt{-Q^2+L \omega^{1/2} } }{L^2 \pi} \;, \qquad 
s = \frac{2 \pi \omega ^{1/2} \sqrt{-Q^2 +L \omega ^{1/2}}}{L^2 \kappa ^2} \;.
\end{align}
To write all the expressions in terms of $Q$ and $\omega$, we use the relation $r_h^2 + Q^2 = \sqrt{\omega} L$.

\subsubsection{On-shell action} \label{OnShellActionAdS5}

Following closely the previous section on AdS$_4$, we evaluate the on-shell action. 
\begin{align} \label{OnShellActionExponentialWAdS5} 
s_{on-shell} &=  \frac{(2-4c_\phi+ \Lambda_\phi) Q^3 - 12 c_F Q^2 \omega^{1/2} L +3 L^2 \omega }{6 L^5 \kappa ^2}\;.
\end{align}
Note that the scalar boundary terms are finite at the boundary.  We do not need to impose a condition on $\Lambda_\phi$ or $c_\phi$, which is different compared to the AdS$_4$ case.  
From this on-shell action, we identify the thermodynamic potential (density)
\begin{align}
G = - \frac{(2-4c_\phi+ \Lambda_\phi) Q^3 - 12 c_F Q^2 \omega^{1/2} L +3 L^2 \omega }{6 L^5 \kappa ^2}\;. 
\end{align}
Again we check that this grand potential is identical to the pressure of the system.  

The corresponding stress energy tensor is given by \eqref{FullStressEnergyTensor}
\begin{align} \label{StressEnergyTensorExponentialWAdS5}
\begin{split}
&E = \langle T_{tt} \rangle = \frac{(4c_\phi-2 - \Lambda_\phi) Q^3 - 12 c_F Q^2 \omega^{1/2} L + 9 L^2 \omega }{6 \kappa^2 L^5 } \;,  \\ 
&P = \langle T_{xx} \rangle = \langle T_{yy} \rangle=  \frac{(2-4c_\phi + \Lambda_\phi) Q^3 - 12 c_F Q^2 \omega^{1/2} L +3 L^2 \omega }{6 \kappa^2 L^5} \;.  
\end{split}
\end{align}
Here $E$ and $P$ are identified as energy and pressure. The pressure is nothing but the grand potential, $P=-G$. 
 
One can also explicitly compute the mass to find from \eqref{MassPressure} and \eqref{MassPressureEquivalence}
\begin{align} \label{MassExponentialWAdS5}
M &= \frac{(4c_\phi-2 - \Lambda_\phi) Q^3 - 12 c_F Q^2 \omega^{1/2} L + 9 L^2 \omega }{6 \kappa^2 L^5 } \;.  
\end{align}
Thus we confirm that the mass and the grand potential both depends on the parameters $\Lambda_\phi, c_\phi$ and $c_F$. One can readily check that the following thermodynamic relation holds 
\begin{align}
\Omega (T,V,\mu) = -P = M - Ts - \mu q \;, \qquad \text{for} \quad c_F=0 \;.
\end{align}

\subsubsection{Grand canonical ensemble $c_F=0$} \label{sec:PolynomialWGrandCanonical}

For the grand canonical ensemble ($c_F=0$), we examine possible quantizations as we have done in the previous section \S \ref{sec:BoundaryProblemExponentialW}. All the possible quantizations are legitimate with general $\Lambda_\phi$ and $c_\phi$. One crucial information on the possible quantization is the falloff of the scalar field given in \eqref{ScalarFalloffBFBound} for mass saturating the BF bound. If $\alpha \neq 0$, we are required to impose the condition \eqref{ConditionExponentialWBFBoundAlpha} or special cases of that. 

For $\alpha=0$, we impose the condition 
\begin{align} \label{conditionAdS5Gubser}
\Lambda_\phi = 4c_\phi -2 \;, 
\end{align} 
that comes from \eqref{ConditionExponentialWBFBoundAlphaZero} for $d=4$ to have a consistent variational problem. Here the scalar is realized as \eqref{AdS5ScalarFalloff} with the faster falloff. This has an important implication. Below this condition is shown to be consistent with the differential form of the thermodynamic first law and also to fix the mass to ADM mass.  

This quantization gives the expectation value for the dual field theory 
\eqref{ExpectationValuesExponeitialWBFBoundAlphaZero}
\begin{align}  \label{ExpectationValueUnfixed}
\langle \mathcal O_{\alpha=0} \rangle =  \frac{c_\phi \beta}{\tilde \kappa^2} =  \sqrt{\frac{2}{3}} \frac{c_\phi Q^2}{\tilde \kappa^2} \;, \qquad 
 \langle \mathcal O_\mu \rangle =  - q \;.
\end{align}
The on-shell action and the grand potential $\Omega$ are 
\begin{align} 
G = -s_{on-shell} &= - \frac{(2-4c_\phi+ \Lambda_\phi) Q^3  +3 L^2 \omega }{6 L^5 \kappa ^2}\;.
\end{align}
Note that we do not impose a condition on $\Lambda_\phi$ or $c_\phi$ yet. 
The stress energy tensor is  
\begin{align} 
\begin{split}
&E =M= \langle T_{tt} \rangle = \frac{(4c_\phi-2 - \Lambda_\phi) Q^3 + 9 L^2 \omega }{6 \kappa^2 L^5 } \;,  \\ 
&P = \langle T_{xx} \rangle = \langle T_{yy} \rangle=  \frac{(2-4c_\phi + \Lambda_\phi) Q^3  +3 L^2 \omega }{6 \kappa^2 L^5} \;.  
\end{split}
\end{align}
Here $E$ and $P$ are energy density and pressure. The pressure is nothing but the grand potential. The grand potential has the relation $G=- P = M - Ts - \mu q $ for general $\Lambda_\phi, c_\phi$. 
 
Upon imposing the differential from of the first law of thermodynamics $ d\Omega = -s dT - q d\mu $, we are required to impose the same condition given in \eqref{conditionAdS5Gubser}. 
Note this condition is the one we have from the consistent variational problem. This happens because the grand potential depends on the parameters $\Lambda_\phi, c_\phi$, while $s, T, q, \mu$ are independent of the parameters similar to AdS$_4$. Then
\begin{align} 
G =  - \frac{3 L^2 \omega }{6 L^5 \kappa ^2}\;,
\end{align}
and 
\begin{align} 
&E =M =  \langle T_{tt} \rangle = \frac{9 L^2 \omega }{6 \kappa^2 L^5 } \;,  \qquad 
P = \langle T_{xx} \rangle = \langle T_{yy} \rangle=  \frac{3 L^2 \omega }{6 \kappa^2 L^5} \;.  
\end{align}

We mention an important implication of the boundary terms. We only fix the combination of $\Lambda_\phi$ and $c_\phi$ through \eqref{conditionAdS5Gubser}. One parameter remains unfixed. Even though the grand potential and stress energy tensors are all fixed, the expectation value of the dual scalar field remains unfixed as can be checked in \eqref{ExpectationValueUnfixed}. We further comment on this below as field theory shares similar properties  \cite{Jackiw:1999qq} (see also \cite{Jackiw:1984zi,Jackiw:1999yp}). 

Previously, three counter terms (different from ours) with undetermined coefficients were considered in the context of linear dilaton gravity \cite{Mann:2009id}. For a certain value of dilaton coupling, one of the coefficient remains unfixed for a well defined variational problem. The resulting on-shell action and conserved charges are shown to be independent of the unfixed coefficient, while the field theory expectation value was not mentioned there. This is similar to our observation done in this section.

\subsubsection{semi canonical ensemble $c_F=1$} \label{sec:canonicalExponentialWAdS5}

The on-shell action and Free energy, after imposing the condition \eqref{conditionAdS5Gubser}, are 
\begin{align}
F = -s_{on-shell} = - \frac{ 3 L^2 \omega- 12 Q^2 \omega^{1/2} L }{6 L^5 \kappa ^2}\;. 
\end{align}
Thus this Helmholtz free energy is a Legendre transformation from the grand potential, $F = \Omega + \mu q = M - T s$. We further confirmed that the differential form of the first law is satisfied, $ d\Omega = -s dT + \mu d q $. 

We comments on some results of the holographic renormalization of the semi canonical ensemble. Energy and pressure are given by 
\begin{align} 
\begin{split}
&E_F = \langle T_{tt} \rangle = \frac{ 9 L^2 \omega - 12 Q^2 \omega^{1/2} L}{6 \kappa^2 L^5 } \;,  \\ 
&P_F = \langle T_{xx} \rangle = \langle T_{yy} \rangle=  \frac{3 L^2 \omega - 12 Q^2 \omega^{1/2} L}{6 \kappa^2 L^5} \;.  
\end{split}
\end{align}
It is curious to notice that the pressure is again the negative of the free energy $P_F = -F$.  One can also explicitly compute the mass, \eqref{MassExponentialW}, to find 
\begin{align}
M_F &= E_F = \frac{9 L^2 \omega - 12 Q^2 \omega^{1/2} L}{6 \kappa^2 L^5 } \;.  
\end{align}
This mass $M_F$ does not play the role of mass in the Helmholtz free energy. See the similar discussion at the end of the section \S \ref{sec:EMDAdS4}.

\subsection{Interpolating solution}  \label{sec:Interpolating}

In this section, we consider the interpolation solution with scaling solution in IR and the AdS$_4$ in UV, which has attracted much attentions recently \cite{Ogawa:2011bz}\cite{Huijse:2011ef}. The action is the same as \eqref{EMDAction} with $d=3$, and we use the coordiantes system considered in \cite{Ogawa:2011bz}  
\begin{align}  \label{InterpolatingSol333}
\begin{split}
ds^2 &=  -\frac{r^2}{L^2}f(r) dt^2 + \frac{r^2}{L^2} dx^2 + \frac{r^2}{L^2} d y^2 + \frac{L^2}{r^2} g(r) dr^2 \;,  \\
g(r) &= \left( 1 + \frac{r_F^{4}}{r^{4}} \right)^{\frac{1}{2}}, \qquad  
f(r)=\frac{k_0 (r/r_F)^{3}}{1+k_0 (r/r_F)^{3}} \;. 
\end{split}
\end{align} 
The solution is specified by 
\begin{align}
\begin{split}
W(r) &=\frac{2 L^6 Q^2 \kappa ^4}{3 } \frac{ \left(k_0 r^3+{r_F}^3\right)^2 \left(r^4+{r_F}^4\right)^{3/2} }{r^6 {r_F}^6 \left(9 r^4+2 k_0 r^3 {r_F}+11 {r_F}^4\right)} \;, \\
V(r) &= -\frac{8 k_0^2 r^8 \left(3 r^4+4 {r_F}^4\right)+7 {r_F}^6 r^2 \left(9 r^4+11 {r_F}^4\right)+2 k_0 r^5 \left(30 r^4 {r_F}^3+41 {r_F}^7\right)}{4 L^2 \left(k_0 r^3+{r_F}^3\right)^2 \left(r^4+{r_F}^4\right)^{3/2}} \;, \\
\phi '(r) &= - \sqrt{\frac{2 {r_F}^3 \left(3 r^4-2 k_0 r^3 {r_F}+{r_F}^4\right)}{r^2 \left(k_0 r^3+{r_F}^3\right) \left(r^4+{r_F}^4\right)}} \;.
\end{split}
\end{align}

The solution reveals that the gauge field $A$ has the following form  
\begin{align} \label{GaugeFieldInterpolatingSol}
A_t =\frac{3 {r_F}^3}{2 k_0 L^4 Q \kappa ^2} - \frac{9 {r_F}^6}{4 \left(k_0^2 L^4 Q \kappa ^2\right) r^3} + \mathcal O(r^{-4}) \;. 
\end{align}
As usual, the constant term is the chemical potential, $\mu = \frac{3 {r_F}^3}{2 k_0 L^4 Q \kappa ^2} $. We compute the charge $q =- \frac{2}{\kappa^2} \sqrt{-g} W(\phi) F^{rt} = Q$. Thus $\mu q = \frac{3 {r_F}^3}{2 k_0 L^4 \kappa ^2} $. Note that the coefficient of the second term in \eqref{GaugeFieldInterpolatingSol} is not directly related to the charge density. This is essential feature due to a nontrivial dilaton coupling. Some fraction of the conserved charge comes from the scalar field through the dilaton coupling $W(\phi)$.  This is contrast to the EMD solutions we consider in \S \ref{sec:EMDAdS4} and \S \ref{sec:EMDAdS5}. They have the property $W \sim \frac{1}{4}$ at the spatial boundary, and thus effectively the charge come from the gauge field alone.  

The mass of the scalar field can be shown to be 
\begin{align}
\begin{split}
m^2_\phi L^2 &= -\frac{9}{4}  \;.
\end{split}
\end{align}
It saturates the BF bound. This is consistent with the radial fall-off of the scalar field 
\begin{align} \label{ScalarInterpolLargeR}
\phi = \frac{2\sqrt{2}}{\sqrt{3 k_0}} \frac{r_F^{3/2}}{r^{3/2}}+ \mathcal O(r^{-5/2}) \;.
\end{align} 
Compared to the general falloff behavior of the scalar, $\phi \to \frac{\alpha }{r^{3/2}} \log r + \frac{\beta }{r^{3/2}}$, this particular solution is realized with the faster falloff. Thus the analysis is similar to the AdS$_5$ done in \S \ref{sec:EMDAdS5}. The variational problem at hand is the case with $\alpha=0$. We come back to details below.

\subsubsection{On-shell action} 

Let us compute the on-shell action and the corresponding stress energy tensor. 
Similar to AdS$_5$, the scalar boundary contributions are finite. This is related to the fact that the mass of the scalar field saturates the BF bound. By keeping $c_\phi, \Lambda_\phi $ explicitly, we get the density 
\begin{align}
s_{on-shell} &= \frac{(9-12 c_\phi+  4 \Lambda_\phi - 9 c_F) {r_F}^3}{6 k_0 L^4 \kappa ^2} \;.
\end{align}
We follow the notations of the previous sections by abusing our notation, even though the concept of the thermodynamic relations are not appropriate for the zero temperature solutions. In particular, the time direction is not compact and infinite. It is a `density' in time coordinate as well. We identify the `potential' as the negative of the on-shell action
\begin{align}
G &= - \frac{(9-12 c_\phi+  4 \Lambda_\phi - 9 c_F) {r_F}^3}{6 k_0 L^4 \kappa ^2} \;.
\end{align}

The corresponding stress energy tensor is given by \eqref{FullStressEnergyTensor}
\begin{align}
\begin{split}
E &= \langle T_{tt} \rangle = \frac{(12 c_\phi -4 \Lambda_\phi - 9 c_F)  {r_F}^3 }{6 k_0 L^4 \kappa ^2 }  \;, \\
P &= \langle T_{xx} \rangle = \langle T_{yy} \rangle = \frac{ (9-12 c_\phi + 4 \Lambda_\phi - 9 c_F) {r_F}^3}{6 k_0 L^4 \kappa ^2 }  \;. 
\end{split}
\end{align}
The physical quantities depend on the boundary contributions through the parameters $c_\phi, \Lambda_\phi$ and $c_F$. Note that the energy would vanish if one would be able to choose $c_\phi = \Lambda_\phi=c_F=0 $, while the pressure would be finite. We see below that it is not the case. Let us consider the $c_F=0$ and $c_F \neq 0$ cases separately.

\subsubsection{Case with $c_F=0$} 

If one chooses $c_F=0$, one works with fixed chemical potential. For $\alpha=0$, the variational problem we considered in \S \ref{sec:PolynomialWBFBound} (and in \S \ref{sec:ExponentialWBFBound} for $c_F=0$), instruct us to choose \eqref{ConditionExponentialWBFBoundAlphaZero} 
\begin{align}  \label{ConditionInterpolating33}
\Lambda_\phi = (2c_\phi -1) \frac{d}{2} = 3c_\phi -3/2 \;. 
\end{align}
The parameters are partially fixed by the consistency condition of the well-defined variational problem. The corresponding expectation values of the dual operators are given in \eqref{ExpectationValuesExponeitialWBFBoundAlphaZero}
\begin{align} 
\langle \mathcal O_{\alpha=0} \rangle =  \frac{c_\phi \beta}{\tilde \kappa^2}= \frac{c_\phi }{\tilde \kappa^2}\frac{2\sqrt{2}}{\sqrt{3 k_0}} r_F^{3/2}\;, \qquad 
 \langle \mathcal O_\mu \rangle =  - q \;.
\end{align}
Thus the expectation value of the scalar in dual field theory depends on the undetermined parameter $c_\phi$. To fix this, it is required to have further input from the field theory side.  

Let us go back to energy and pressure for $c_F=0$. After using $\Lambda_\phi = 3c_\phi -3/2$ given in \eqref{ConditionInterpolating33}, we get 
\begin{align}
E &=  \frac{ {r_F}^3 }{ k_0 L^4 \kappa ^2 }  \;, \qquad
G = -P = - \frac{ {r_F}^3}{2 k_0 L^4 \kappa ^2 }  \;. 
\end{align}
It is interesting to observe that the `potential' $G$ satisfies a relation even at zero temperature 
\begin{align}
G = -P= E - \mu q \;.
\end{align}
It resembles that of thermodynamics without the term $Ts$.   
We also note that the energy and pressure satisfy the traceless condition 
\begin{align}
\langle T^\mu_{\ \mu} \rangle = -E + 2 P = 0 \;.
\end{align}
Apparently the interpolating solution still respect the conformal invariance even though the interior is much modified with the hyperscaling violation geometry.

\subsubsection{Case with $c_F \neq 0$} 

Let us briefly mention on $c_F \neq 0$. This case requires the sub-leading part of the gauge field in the analysis of the variational problem. It is different from fixing the conserved charge because the scalar coupling also contributes to the charge. 

For $\alpha=0$, the general variational problem is analyzed in \S \ref{sec:PolynomialWBFBound}. In particular, the mixed boundary condition is given by \eqref{MixedBCPolynomialWBFBound}
\begin{align}   
(c_F -1) \delta \log \mu + c_F  \delta \log Q_F 
=  \delta \log \left( \mu^{3c_F -1} q^{c_F}\right)  =0 \;,
\end{align}
where the solution gives $Q_F =3 (\mu q) \mu L^4 \kappa^2$. 
The corresponding vacuum expectation value of the dual scalar operator is 
\begin{align}
\langle \mathcal O_{\alpha=0} \rangle = -\frac{ c_\phi }{\tilde \kappa^2 } \beta = -\frac{c_\phi }{\tilde \kappa^2}\frac{2\sqrt{2}}{\sqrt{3 k_0}} r_F^{3/2}\;,
\end{align} 
where we use $\beta = \frac{2\sqrt{2}}{\sqrt{3 k_0}} r_F^{3/2} $.
The expectation value depends on the parameter $c_\phi$. Moreover, the variational problem requires the following condition given in \eqref{ConditionMixedBCPolynomialWBFBound}. 
\begin{align} \label{Condition333}
\left( \Lambda_\phi+[1-2c_\phi] \frac{d}{2} \right) \beta + \frac{c_F}{c_W} k \tilde \kappa^2 \mu q  = 0\;. 
\end{align}
For the particular solution \eqref{InterpolatingSol333}, satisfying the condition \eqref{Condition333} amounts to 
\begin{align} \label{ConditionOnParametersInterpolatingNonZerocF}
\Lambda_\phi = 3c_\phi - \frac{3}{2} +\frac{3}{2} c_F  \;, 
\end{align}
where we use $d=3, k=-4/3$, $c_W =\frac{2\sqrt{2}}{\sqrt{3} k_0} r_F^{3/2} $. Note also this condition reduces to the previous case \eqref{ConditionInterpolating33} for $c_F=0$. 

To consider the `semi canonical' ensemble, we add the boundary term \eqref{MaxwellBoundaryTerm}. The on-shell action and `Free energy' are modified, after imposing the condition \eqref{ConditionOnParametersInterpolatingNonZerocF}, to  
\begin{align}
F = -s_{on-shell} =  \frac{(c_F-1) {r_F}^3}{2 k_0 L^4 \kappa ^2}  \;.
\end{align}
Here again, we abuse our notation for the `free energy.' 
The corresponding stress energy tensor is given by 
\begin{align}
E &= 
\frac{(2- 5 c_F)  {r_F}^3 }{2 k_0 L^4 \kappa ^2 } \;, \qquad
P = 
\frac{ (1-c_F) {r_F}^3}{2 k_0 L^4 \kappa ^2 }  \;. 
\end{align}
These are functions of $c_F$. Setting $c_F=1$ does not correspond to fixing charge, the coefficient $c_F$ can be considered on equal footing as the other two coefficients $\Lambda_\phi$ and $ c_\phi$.

\section{Two physical implications} 

Here we consider two physical implications that manifest themselves throughout the paper. 

\subsection{Finite boundary terms} \label{sec:FiniteCounterTerms}

In this section we have a fresh look for the finite boundary or counter terms we encountered above. In particular, we compare the situation to that of the field theory side.  

Renormalization in Quantum field theory is a way to render divergent physical quantities, such as mass and coupling constants, into finite ones by subtracting the divergence using the radiative corrections. Analogous to the gravity theories we considered above, there have been cases when the radiative corrections of field theories are undetermined, not to mention finite. This phenomena has been investigated by Roman Jackiw \cite{Jackiw:1999qq} (see also \cite{Jackiw:1984zi} \cite{Jackiw:1999yp}).

Jackiw asked the following question. {\it Can one formulate a criterion that will settle a priori whether the radiative correction produces a definite or indefinite result?} In \cite{Jackiw:1999qq}, the basic rule of thumb has been provided. If the computed radiative correction, when inserted into the bare Lagrangian, preserves the renormalizability and retains the symmetries of the theory, the radiative correction does not produce a definite result. Three different classes of examples are presented in \cite{Jackiw:1999qq}: (a) radiative corrections are uniquely determined by spoiling renormalizability or gauge invariance such as $g\!-\!2$ Pauli term in QED or photon mass in Schwinger model, (b) radiative corrections are not determined because, due to chiral anomaly, there is no symmetry prohibiting to insert photon mass in the chiral Schwinger model, and (c) Radiative correction can be determined depending on which symmetry, vector or axial vector, we choose to preserve in the case of triangular graph of axial vector anomaly. Thus for the cases (b) and (c), further `experimental' inputs are necessary to fix the radiative corrections. See more details in \cite{Jackiw:1999qq}.  
We examine the available examples in holography and compare them with those of the examples in field theory.  

\subsubsection{Examples in Holography}

Here we consider several known examples in holography that claim to have finite counter terms as well as some relevant cases that provide close connection to field theory radiative corrections. The holographic boundary terms are required by well defined variational problems. The counter terms are required to yield the finite on-shell action and stress energy tensor. Sometimes the boundary terms are also used to cancel the divergences, and thus we put them in equal footing. They are far from unique, even though constructed from the local and covariant functions of the intrinsic boundary geometry, such as the induced metric $\gamma_{ij} $ and the Ricci scalar and tensor built from $\gamma, \mathcal R^{(d)}$. This is true even for the simplest geometry. For example, AdS$_3$ requires only a term $S_{ct} = -\frac{1}{\ell} \int \sqrt{-\gamma}$ to cancel the divergence in the stress energy tensor. In general, one can add the terms such as $c_n \mathcal R^n$, $n \geq 1$. The coefficients $c_n$ are not determined because these terms vanish too fast to contribute to the finite part of the stress tensor. Let us recount a few examples.

\paragraph{\it Anomalies}
In \cite{Henningson:1998gx}\cite{Balasubramanian:1999re}, 
holographic renormalization for the AdS has been carried out to match the field theory expectations 
for various physical quantities including anomalies.

For the metric of the pure AdS$_3$ 
\begin{align}
ds^2 = \frac{L^2}{r^2} dr^2 + \gamma_{ij} dx^i dx^j \;,
\end{align}
one can compute the extrinsic curvature to be 
\begin{align}
\Theta_{ij} = -\frac{r}{2\ell} \partial_r \gamma_{ij} \;, \qquad 
\Theta = -\frac{r}{2\ell} \gamma^{ij} \partial_r \gamma_{ij} \;.
\end{align}
To evaluate this we use the expansion of the metric 
\begin{align}
\gamma_{\mu\nu} = r^2 \gamma_{\mu\nu}^{(0)}  + \gamma_{\mu\nu}^{(2)} + \cdots \;, \qquad 
\gamma^{\mu\nu} = r^{-2} \gamma^{(0) \mu\nu}  +r^{-4} \gamma^{(2) \mu\nu} + \cdots \;. 
\end{align} 
Then by taking the trace of the stress energy tensor for pure AdS$_3$, we get  
\begin{align}
T^i_{\ i} &= -\frac{1}{8\pi G_{3}} \left( \Theta +\frac{2}{\ell}  \right)
= -\frac{1}{8\pi G_{d+1}}  \frac{1}{\ell r^2} \gamma^{(2) \mu\nu} \gamma_{\mu\nu}^{(0)} = -\frac{\ell}{16 \pi G_{d+1}}  \mathcal R \;,
\end{align}
where boundary limit is taken and $ \gamma^{(2) \mu\nu} \gamma_{\mu\nu}^{(0)} = \frac{\ell^2 r^2}{2} \mathcal R$. 

This conformal anomaly is a consequence of breaking conformal invariance in the process of 
regularization and renormalization. Due to the divergences of the effective action and of the 
boundary extrinsic curvature term, they should be regularized in a way to preserve the 
general covariance, analogue of the gauge invariance of the quantum field theory. 
The regularization procedure picks up a particular representation of the conformal class of the 
boundary theory. In this way, conformal invariance is explicitly broken, and the trace of the stress energy tensor is expected to have a unique answer after the renormalization. 
Compared to the field theory expectations, the features we describe here are similar to the case (a) rather than (b) described above \cite{Jackiw:1999qq}. The holographic renormalization process spoils the general covariance to fix the coefficient of the boundary counter terms.

\paragraph{\it Casimir energy} 
Similar to the anomalies computed and determined above, some holographic examples provide a definite answers for Casimir energy in global AdS. We consider AdS$_5$ in global coordinate with $(t, \theta, \phi, \psi, r)$
\begin{align} 
ds^2 &= - \left(1+\frac{r^2}{L^2} \right) dt^2 + \left(1+ \frac{r^2}{L^2} \right)^{-1}  dr^2 + r^2 d \Omega_{3}^2  \;, 
\end{align} 
where $d \Omega_{3}^2 $ is the metric for $3$ dimensional sphere. The energy density by including the standard counter terms \cite{Balasubramanian:1999re} gives
\begin{align}
T_{tt} &= \frac{3 L}{8 \kappa^2 r^2} + \cdots \;.
\end{align}
We use the following metric of the field theory living at $r\to \infty$ 
\begin{align} 
ds_{FT}^2 = - dt^2 + L^2 \left(d\theta^2 + \sin^2\theta d\phi^2  + \cos^2 \theta d\psi^2\right) \;. 
\end{align}
The field theory energy density is given by (see \S \ref{sec:OnShellAction})
\begin{align}
E = \langle T_{tt} \rangle &= \frac{3 }{8 \kappa^2 L}  \;. 
\end{align}
This is identified as a Casimir energy in the field theory side \cite{Balasubramanian:1999re}. Similar computations give the energy density for AdS$_3$ in global coordinate  
\begin{align}
E &= \langle T_{tt} \rangle = - \frac{1}{\kappa^2 L}  \;.   
\end{align}
See also \cite{Horowitz:1998ha}\cite{Myers:1999psa}\cite{Emparan:1999pm} for more general discussion of Casimir energies in the holographic context.

\paragraph{\it Holographic Examples with finite boundary terms} 
Here we briefly mention some examples of the {\it finite} boundary or counter terms in holography that have been discussed in the literature. 

One class of the examples that has introduced the finite counter term in holographic setup is in the context of R-charged black holes \cite{Liu:2004it}\cite{Kim:2010tf}. The black hole solutions have gauge fields and scalar fields. There the finite counter term, $\phi^2$, was introduced to match the expected mass in AdS$_5$ \cite{Liu:2004it}. In other wards, the first law of thermodynamics requires to have the counter term $\phi^2$, which is actually finite. Similarly, the thermodynamic properties of the R-charged black holes have been analyzed in the context of Sch\"odinger space-time \cite{Kim:2010tf}, where again the finite counter term proportional to $\phi^2$ is required to satisfy the first law.  

Another class of finite counter terms has been considered in holographic studies of thermal and electric responses in \cite{Amoretti:2014zha}\cite{Forcella:2014dwa}. It has been argued that finite counter terms are necessary to yield a consistent physical picture that matches with field theory expectations. 

These two classes of holographic examples are similar to the field theory finite radiative corrections. The finite counter terms are undetermined until `experimental' data are provided. Here the expectations from field theory, such as first law of thermodynamics or transport properties, are required to determine the finite counter terms.

The third class of examples of finite counter term has been related to the probe brane wrapping around $AdS_5 \times S^3$ in the D3-D7 brane system \cite{Karch:2005ms}. It is noted that a new type of finite term is possible. The finite counter term is used to set the on-shell action to vanish for super-symmetric theory. This is an example that the finite term can be fixed by demanding the symmetry discussed in the case (c).

\subsubsection{EMD theories} 

Here we summarize the results of the holographic renormailzation of the EMD theories and compare them with the situations of the field theory.
 
We start with the boundary term for the gauge field, which is describe by the term. 
\begin{align}
S_{b}^F &= \frac{1}{\kappa^2} \int_{ \partial M} d^d x \sqrt{-\gamma} ~2 c_F W n_r F^{rt} A_t \;.
\end{align} 
This is actually the best known finite boundary term. If we do not decide the field theory physical systems, say grand canonical ensemble or semi canonical ensemble, the parameter $c_F$ remains unfixed. This is even more clear with the $W(\phi)$ that bring out the mixing between the gauge and scalar variations. As one can check, the boundary term associated with $c_F$ is always finite because it is nothing but the conserved charge \eqref{ConservedMaxwellCharge}. 
This boundary finite counter term is not directly related to any finite radiative corrections  discussed in \cite{Jackiw:1999qq}. 

There are two additional boundary terms for the scalar fields 
\begin{align} \label{ScalarBouondaryTerms} 
S_{b}^\phi &= \frac{1}{\kappa^2} \int_{ \partial M} d^d x \sqrt{-\gamma} 
\left[ \frac{c_\phi}{2} \phi n^r \partial_r \phi + \frac{\Lambda_\phi}{4 L} \phi^2 \right] \;.
\end{align}
These two terms have the same boundary falloffs in the asymototic AdS boundary because $n^r \partial_r $ is independent of $r$. We consider the general $\Lambda_\phi$ and $c_\phi$ until we are forced to fix them. Choosing $c_\phi=0$, for example, from the beginning could lead a consistent theory in the end, but we do not take that approach. 

As we demonstrate above, the variational problem at the boundary provides certain conditions for the parameters such that the theory is well defined. It turns out that this condition is, in general, different from the conditions that we can get by requiring finiteness of the on-shell action and stress energy tensor. Theses two conditions coincide if the boundary terms are used to cancel divergent contributions. They are in general different when the boundary terms are finite. In that case, the parameters of the theory are not completely fixed even after imposing the first law of thermodynamics. We encounter numerous examples from the consistency of the general variational problem. 

Let us consider the three examples we have considered one by one. 
\begin{itemize}
\item AdS$_4$: The boundary contribution of the scalar kinetic term has a divergent contribution. It is required to cancel the divergence with a condition, $\Lambda_\phi =2 c_\phi-1$ for $c_F=0$, from the boundary counter terms \eqref{ScalarBouondaryTerms}. And then one of the remaining parameter can be fixed by requiring the differential form of the first law of thermodynamics, which we consider as an independent input. Thus the EMD theory with AdS$_4$ asymptotics itself can completely determine the parameters with the thermodynamic first law.    

\item AdS$_5$: The scalar mass saturates the BF bound, and the boundary contribution of the scalar kinetic term is actually finite. Nevertheless, a condition is necessary to have a well defined boundary variational problem $\Lambda_\phi =4 c_\phi-2$ for $c_F=0$. Surprisingly, this particular condition coincides with the requirement to satisfy the differential form of the first law of thermodynamics! The thermodynamic potential, energy and pressure are determined with the condition. Yet, the expectation value of the operator dual to the scalar is a function of a parameter, say $c_\phi$. Thus the physical quantities are not completely fixed even after imposing the first law. 

\item Interpolating solution: The example gives the conserved charge different from the charge provided by the gauge field. Part of the conserved charge comes from the dilaton coupling, and thus from the scalar. The scalar mass saturates the BF bound. The features of the holographic renormalization are similar to the AdS$_5$ example. General variational problem provides a condition $\Lambda_\phi =3 c_\phi-3/2 + 3/2 c_F$. For $c_F=0$, the energy and pressure are fixed and independent of the parameters, while the expectation value of the scalar depends on the parameter $c_\phi$. Thus the boundary terms are not completely determined by the gravity theory. It requires further input or experimental data from the boundary field theory. 
    
\end{itemize}

\subsection{Non-Fermi liquids \& Charge splitting} \label{sec:HiddenCharge}

In this last section we examine the conserved charge from the gauge field point of view. We focus on the case with asymptotic AdS boundary. Using the dominant profile of the gauge field at the boundary $r=\infty$ used in \eqref{FieldStrengthBoundaryQ}
\begin{align}
F_{rt} = \frac{Q_F}{r^{\lambda_Q}} \;,
\end{align}
where $\lambda_Q$ and $Q_F$ depend on the details of the solution. The corresponding conserved charge \eqref{ConservedMaxwellCharge} at the boundary can be expressed in a slightly different form 
\begin{align} 
q = -\frac{4}{2\kappa^2} \sqrt{-g} W F^{rt} 
= \frac{4}{2\kappa^2 L^{d-1}} \frac{Q_F}{r^{\lambda_Q + 1 - d}} W(\phi) \;. 
\end{align}
Because the charge $q$ is conserved, there are tension between the field strength $F_{rt} $ and $W(\phi)$. We go back to our examples to see the qualitatively different physics. 

Let us consider the AdS$_5$ example in \S \ref{sec:EMDAdS5}, it is easy to see that the coupling $W(\phi)$ and gauge field behave, at the boundary, as 
\begin{align}
W = \frac{1}{4} \;,  \qquad F_{rt} = \frac{2\sqrt{2\omega}Q}{r^3} \;,
\end{align} 
where $ \lambda_Q =3, Q_F = 2\sqrt{2\omega}Q$. 
Thus the conserved charge for AdS$_5$ (and thus $d=4$) is  
\begin{align} 
q = \frac{\sqrt{2\omega}Q}{\kappa^2 L^{3}} \;. 
\end{align} 
Thus we check that the contribution to the conserved charge entirely come from the gauge field in this particular solution. It is also straightforward to see the same for the AdS$_4$ solution in \S \ref{sec:EMDAdS4}.

These examples provide the physical properties that resembles the Fermi liquid state of matter at low temperature $T$ \cite{Gubser:2009qt}. For example, entropy density $s$ is proportional to the temperature. Similarly, the specific heats $C$ at constant charge density and constant chemical potential are also coincide with the entropy density  
\begin{align}
s \sim C \sim T \;.
\end{align} 

Let us turn to the interpolating solution we consider in \S \ref{sec:Interpolating}. There we see that 
\begin{align} \label{WAndFInterpolating}
W = \frac{2 k_0^2 L^6 Q^2 \kappa ^4}{27 r_F^6} r^2 \propto \phi^{-4/3}\;,  \qquad F_{rt} = \frac{27 r_F^6}{4 \left(k_0^2 L^4 Q \kappa ^2\right) } \frac{1}{r^4} \;,
\end{align} 
which give the conserved charge $q = Q$. This example is very different from the previous examples. In particular, the coupling has a rather non-trivial behavior at the boundary. (The physical coupling, $e$ in $W \sim \frac{1}{e^2}$, behaves as $e \to 0$ at the boundary.) It actually dominates at large radius compared to the field strength. This means that the conserved charge does not entirely come from the gauge field. In a way, the charge is hidden in the geometry because the dilaton field can be considered as a part of the geometry from the string theory point of view. It will be interesting to closely examine whether this charge splitting has some direct or indirect connections to the fractionalization of charge that has been considered in the holographic context \cite{Sachdev:2010um}\cite{Hartnoll:2011pp}\cite{Huijse:2011ef}. 

Due to this interesting competition between the gauge and scalar fields in contributing to the conserved charge, the model provides a highly non-trivial and interesting physics. For example, it provides an example of non-Fermi liquid state of matter. Physical properties are much deviated from the conventional ones. The interpolating solution we consider in \S \ref{sec:Interpolating} reveals 
\begin{align}
s \sim C \sim T^{4/7} \;, 
\end{align}
where $s$ and $C$ are the entropy density and specific heat \cite{Ogawa:2011bz}. They are readily different from the Fermi liquid case $s \sim C \sim T$.

In closing, we mention that the dilaton coupling and a scalar potential in general EMD theories can be used to accommodate two independent parameters (dynamical and Hyperscaling violation exponents), which are directly and/or indirectly responsible for various interesting physical properties. For example, the existence of the exotic phases of matter \cite{Dong:2012se}\cite{Kim:2012nb} can be signified by the holographic entanglement entropy \cite{Ryu:2006bv}\cite{Ryu:2006ef}\cite{Nishioka:2009un}. For certain parameter ranges of the dynamical and Hyperscaling violation exponents, the holographic entanglement entropy interpolates between the logarithmic violation and extensive volume dependence of entanglement entropy. The former has been advertised to indicate the presence of the Fermi surfaces \cite{Ogawa:2011bz}\cite{Huijse:2011ef} (see also \cite{Hartnoll:2012wm}).

\section{Conclusion} \label{sec:Conclusion}

We have examined the holographic renormalization focusing on the role of the dilaton coupling. Due to this coupling, we consider the boundary variational problem that includes the boundary terms for the gauge and scalar fields together in \eqref{GeneralBoundaryTerms}. There the parameters $\Lambda_\phi, c_\phi$ are introduced for the scalar, and $c_F$ for the gauge field. The corresponding on-shell action and stress energy tensor are determined in \eqref{FullOnShellAction} and \eqref{FullStressEnergyTensor}. We also check that the mass density and pressure evaluated by the Brown-York formula \cite{Brown:1992br} are equivalently given by the components of the field theory stress energy tensor \eqref{MassPressureEquivalence}.    

From the analysis of the boundary value problem, we conclude that the mixed boundary condition between the gauge and scalar fields are indeed possible. Let us illustrate this with a simple example for a scalar with $\alpha =0$ in the boundary expansion of the massless scalar $\phi \to \alpha  + \frac{\beta }{r^{d}} $, along with the gauge field \eqref{FieldStrengthBoundaryQ} in the context of exponential coupling $W \sim e^\phi$ given in \S \ref{sec:BoundaryProblemExponentialW}. We have the general variation for $\alpha=0$ as \eqref{AlphaZeroVariation}
\begin{align}  
\left[\frac{(\Lambda_\phi -(c_\phi -1) d) \beta}{\tilde \kappa^2}   + c_F \mu q c_W \right] \delta \alpha   + (c_F -1) \mu q \frac{\delta \mu}{\mu} + c_F \mu q \frac{\delta Q_F}{Q_F} =0 \;.
\end{align} 
We impose the condition $Q_F = const. $ for $c_F=1$ that gives the expectation value for the dual current operator as $\langle \mathcal O_{Q_F} \rangle =  \frac{\mu q}{Q_F}$ as in \eqref{AlphaZeroVEV}. Now the expectation value of the dual scalar operator is given by the expectation value of the dual current operator \eqref{AlphaZeroVEV}
\begin{align} 
\langle \mathcal O_{\alpha=0} \rangle = \frac{(\Lambda_\phi -(c_\phi -1) d) }{\tilde \kappa^2} \beta  +  c_W  \langle \mathcal O_{Q_F} \rangle Q_F \;.
\end{align}
This example demonstrates several important results we have advertised throughout the paper. 

\begin{itemize}
\item First, the expectation value of a scalar is not only a function of $\beta$, but also the expectation value of the dual current operator $\langle \mathcal O_{Q_F} \rangle $. 

\item $Q_F$ is not identical to $q$, which is conserved charge. This happens because of the dilaton coupling. This demonstrates that the fixed charge ensemble $q$ is not coincide with the fixed $Q_F$ ensemble.  

\item The expectation value of the dual scalar actually depends on the parameters $\Lambda_\phi$ and $c_\phi$. This demonstrates the finite boundary terms that are prevalent for the theories with a scalar field. The finite boundary or counter terms are examined along with previous examples in the literature in \S \ref{sec:FiniteCounterTerms}.  
 
\end{itemize}
 
We examine the program with three EMD solutions with asymptotic AdS boundary. Depending on the particular solutions, we demonstrate the consistency between the boundary variational problem and the on-shell action. Let us take the example of AdS$_5$ in \S \ref{sec:EMDAdS5} to summarize. The gauge field $A$ at the boundary is given by 
\begin{align}
A_t = \frac{ Q \sqrt{2}}{L} - \frac{ Q \sqrt{2\omega }}{r^2}  \;. 
\end{align}
The two terms are directly related to the chemical potential $\mu = \frac{ Q \sqrt{2}}{L} $ and the charge density $q = \frac{1}{\kappa ^2}\frac{Q\sqrt{2 \omega }}{L^3}$. This happens because $W \sim 1/4$ at the boundary. The same property is shared by the example given in \S \ref{sec:EMDAdS4} as well. The corresponding physical properties are similar to those of the Fermi liquid states in \S \ref{sec:HiddenCharge}. This is contrasted to the other example in \S \ref{sec:Interpolating} that demonstrates non-Fermi liquid states. It is related to the non-trivial boundary profile of $W$ given in \eqref{WAndFInterpolating}.  
The scalar field saturates the BF bound. 
\begin{align}
\phi = \sqrt{\frac{2}{3}} \frac{Q^2}{r^2}  \;.
\end{align}
This solution is realized with the faster falloff of the scalar with $\alpha=0$ in $\phi \to \frac{\alpha }{r^2} \log r + \frac{\beta }{r^2}$.

The on-shell action and the Grand potential have been evaluated in \S \ref{sec:PolynomialWGrandCanonical}
\begin{align} 
G = -s_{on-shell} &= - \frac{(2-4c_\phi+ \Lambda_\phi) Q^3  +3 L^2 \omega }{6 L^5 \kappa ^2}\;.
\end{align}
They are functions of the two parameters $c_\phi$ and $ \Lambda_\phi$. These coefficients are partially fixed by imposing the differential form of the first law of thermodynamics $d\Omega = -s dT - q d\mu$ to satisfy 
\begin{align} 
\Lambda_\phi = 4c_\phi -2 \;, 
\end{align} 
as in \eqref{conditionAdS5Gubser}. This condition turns out to be the same condition \eqref{ConditionExponentialWBFBoundAlphaZero} required by the consistent variational problem for $\alpha=0$. Then the mass evaluated by stress energy tensor agrees with the ADM mass. 

While all the thermodynamic quantities are fixed with this condition, the expectation value of the dual scalar operator  depends on the parameter $c_\phi$ in \eqref{ExpectationValueUnfixed}
\begin{align}  
\langle \mathcal O_{\alpha=0} \rangle =  \frac{c_\phi \beta}{\tilde \kappa^2} =  \sqrt{\frac{2}{3}} \frac{c_\phi Q^2}{\tilde \kappa^2} \;.
\end{align}
It is an example of finite boundary term. To fix the expectation value, further information is required from the field theory side. Finite counter terms have been considered previously and briefly summarized in \S \ref{sec:FiniteCounterTerms}. \\

In this paper, we have explored general possibilities to have a mixed boundary condition between the scalar and gauge fields. For example, we have observed the expectation value of the dual scalar field can be a function of the expectation value of the current operator. It will be interesting to find their applications. 

It will be also interesting to generalize this program to the backgrounds with different asymptotic boundaries such as Lifshitz space or Schr\"odinger space with emphasis on the dilaton coupling. Holographic renormalization has been done successfully in \cite{Charmousis:2010zz} in the context of the EMD theories with Lifshiz asymptotics.

\section*{Acknowledgments}

We are very grateful to Sumit Das, Ori Ganor, Elias Kiritsis, Robert Mann and Ioannis Papadimitriou for illuminating discussions, correspondences and valuable comments on the draft. 
This work is partially supported by NSF Grant PHY-1214341.

\end{document}